\documentclass[12pt]{article}

\usepackage{amssymb}
\usepackage{color}
\usepackage{epsfig,amssymb,amsfonts,amsmath,graphicx,dsfont,cite,xfrac}
\usepackage{authblk}
\usepackage{subcaption}

\usepackage[colorlinks=true,linkcolor=blue,citecolor=blue]{hyperref}

\newcommand{\be}{\begin{equation}}
\newcommand{\ee}{\end{equation}}
\newcommand{\bea}{\begin{eqnarray}}
\newcommand{\eea}{\end{eqnarray}}

\title{SU(1,1) and SU(2) Approaches to the Radial Oscillator: Generalized Coherent States\\ 
and Squeezing of Variances}

\author[1]{Oscar Rosas-Ortiz}
\author[2]{Sara Cruz~y~Cruz}
\author[1,2]{Marco Enr\'{\i}quez}

\affil[1]{\footnotesize Physics Department, Cinvestav, AP 14-740, 07000
M\'exico DF, Mexico}

\affil[2]{\footnotesize UPIITA, Instituto Polit\'ecnico Nacional, Av. IPN 2580, C.P. 07340, M\'exico DF, Mexico}

\date{}
\begin{document}

\maketitle

\begin{abstract}
It is shown that each one of the Lie algebras $su(1,1)$ and $su(2)$ determine the spectrum of the radial oscillator. States that share the same orbital angular momentum are used to construct the representation spaces of the non-compact Lie group $SU(1,1)$. In addition, three different forms of obtaining the representation spaces of the compact Lie group $SU(2)$ are introduced, they are based on the accidental degeneracies associated with the spherical symmetry of the system as well as on the selection rules that govern the transitions between different energy levels. In all cases the corresponding generalized coherent states are constructed and the conditions to squeeze the involved quadratures are analyzed. 
\end{abstract}


\section{Introduction}

It is well known that the Lie groups admitted by the mathematical models of physical  phenomena determine the special functions that describe the corresponding quantum states \cite{Mil68, Bar77, Per86}. Of special interest, the Lie groups $SU(1,1)$ and $SU(2)$ play an important role in the study of exactly solvable models like the $n$-dimensional oscillator, Morse oscillator, Coulomb potential, angular momentum, qubits and qudits. 

A remarkable example is given by the one-dimensional oscillator $H=N + \tfrac12$, for which the operators $J_3= \tfrac12 H$ and $J_{\pm} = \tfrac12 (a^{\pm})^2$, with $[a^-, a^+]=\mathbb I$ and $N=a^+ a^-$, satisfy the commutation relations of the Lie algebra $su(1,1)$ \cite{Bar77}. As the Casimir operator is proportional to the identity $C= -\tfrac{3}{16} \mathbb I= \kappa (\kappa -1) \mathbb I$, the Bargmann index $\kappa$ defines two classes of energy eigenstates: even for $\kappa = \tfrac14$ and odd for $\kappa = \tfrac34$. Therefore, the full space of states is covered by two infinite representations of $SU(1,1)$. Another important example, dealing with the subject of the present work, is the 3D~isotropic oscillator. In this case the role of dynamical group is played by the symplectic group $S_p(6) \supset S_p(2) \otimes O(3)$ \cite{Mos71}, where the rotation group $O(3)$ characterizes the quantum states and $S_p(2) \sim SU(1,1)$ generates the admisible energy spectrum. 

Below, we study the radial part of the 3D isotropic oscillator in spherical coordinates and present a development of the related spectrum generating algebras that is based on the factorization method \cite{Dir35,Sch40,Inf51} (see also the review \cite{Mie04}). We find that, in addition to $SU(1,1)$, also $SU(2)$ generates the spectrum of the radial oscillator; a result that has been unnoticed in the literature on the matter until the present work. The presence of $SU(2)$ in our approach is associated with the accidental degeneracies that arise because the spherical symmetry of the system and to the selection rules that govern the transitions between energy levels.

In this work we factorize the radial part of the spherical oscillator Hamiltonian in four different forms and show that the factors can be used to construct the basis elements of the Lie algebras of $SU(1,1)$ and $SU(2)$. Our identification of $su(1,1)$ as the  generating algebra of the radial oscillator shows that the factorization method developed here is equivalent to the representation theory reported in, e.g., \cite{Mos71}. Quite interestingly, we find that the factorization constant $\epsilon_{\ell}$ defining the building-blocks of the $su(1,1)$ algebra corresponds to the Bargmann index $\kappa= \tfrac14 \epsilon_{\ell} = \tfrac{\ell}{2}+ \tfrac34$, which characterizes the representation (here $\ell$ denotes the orbital quantum number). In this form, the full state space is covered by a denumerable set of infinite-dimensional representations of $SU(1,1)$, each one spanned by radial-oscillator states of definite orbital angular momentum. 

On the other hand, the finite-dimensional representations of $SU(2)$ are achieved here in three different forms. The first one is by considering the accidental degeneracy of the oscillator energy eigenvalues. The full space of states is covered in this case by a denumerable set of degeneracy subspaces, each one a representation space of $SU(2)$. A second way is based on the transitions between the states that integrate a given degeneracy subspace. By necessity, the allowed transitions require intermediary states that are not in such a subspace. All these states, the intermediary ones and those belonging to the degeneracy subspaces, give rise to finite representations of $SU(2)$ that are different from the ones indicated above. The third way is in terms of the state vectors that satisfy the rule $s+ \ell = 2j_C$, with $s$ the radial quantum number and $j_C$ a constant that defines the dimension of the representation. In the last two cases the space of states is also decomposed into the direct sums of the corresponding finite-dimensional subspaces.

Provided the above representations we construct different sets of generalized coherent states for the radial oscillator. As far as we know, there is not any previously reported  construction of $SU(2)$ coherent states (also called {\em spin} coherent states or {\em Bloch} states) for this system. Our $SU(2)$ coherent states are either linear combinations of state vectors in a given degeneracy subspace, or linear combinations of states that obey the rule $s+ \ell = 2j_C$. On the other hand, for the $SU(1,1)$ coherent states of the radial oscillator, some results were already obtained in e.g. \cite{Ger88} and \cite{ Aga95}. These are recovered as particular cases in our model at the time that some of the aspects that were unclosed in \cite{Ger88,Aga95} are now clarified in terms of the factorization method. For instance, we show that the generators of the $su(1,1)$ algebra can be written in the two-boson representation of Schwinger \cite{Sch52}, so that the related coherent states are affected by the two-mode squeezing of the quadratures. 

The paper is organized as follows. In Section~\ref{secfac} we provide the basic set of operators that are necessary in the construction of the generating algebras of our approach. Besides, we classify the state space in hierarchies, these are defined as sets of vectors that describe systems with either definite energy or definite orbital angular momentum. Section~\ref{intertwining} includes the construction of the corresponding generating algebras, representation spaces and coherent states. The results for $SU(1,1)$ are reported in Section~\ref{algdm} and the ones for $SU(2)$ in Section~\ref{algde}. In Section~\ref{secdiagonal} we discuss the third way of getting a finite representation of $SU(2)$ and comment some of the generalities of the corresponding coherent states. In Section~\ref{conclu} some conclusions and perspectives of the present work are given. The paper concludes with an appendix containing complementary information that is relevant in different parts of the manuscript but can be consulted in separate form.

\section{Factorization method}
\label{secfac}

In suitable units, the radial Hamiltonian of the spherical oscillator is of the form
\be
H_{\ell} =-\frac{d^2}{dr^2} + \frac{\ell (\ell+1)}{r^2} +\lambda^2 r^2 \equiv -\frac{d^2}{dr^2} + V_{\ell}(r), \qquad \lambda = \frac{m\omega}{\hbar}.
\label{f1}
\ee
The related eigenvalue equation can be written as either 
\be
H_{\ell} \vert n, \ell \rangle_e = E_n \vert n, \ell \rangle_e \quad \mathrm{or} \quad H_{\ell} \vert s, \ell \rangle = E_{s,\ell} \vert s, \ell \rangle,
\label{f3}
\ee
with $E_n = 2\lambda \mathtt{E}_n =\lambda(4s+2\ell +3) \equiv E_{s,\ell}$. Here 
\be
\mathtt{E}_n= n+\tfrac32
\label{f4}
\ee
stands for the dimensionless energy eigenvalue of the spherical oscillator (see Appendix~\ref{ApA}). The corresponding eigenvectors $\vert n, \ell \rangle_e = \vert 2s+\ell,\ell \rangle_e \equiv \vert s, \ell \rangle$ are orthonormal 
\[
{}_e\langle n,\ell \vert n',\ell'\rangle_e = \delta_{nn'} \delta_{\ell \ell'}, \qquad \langle s,\ell \vert s',\ell'\rangle = \delta_{ss'} \delta_{\ell \ell'},
\]
 while the {\em principal}, {\em orbital} and {\em radial} quantum numbers, $n$, $\ell$ and $s$ respectively, satisfy the condition 
\be
n= 2s +\ell, \qquad n,s,\ell=0,1,\ldots
\label{f2}
\ee
This last indicates {\em accidental degeneracies} $d=\mbox{deg}(E_n)$ of the energy eigenvalue ${E}_n$. Namely, ${E}_n$ is $(\frac{n}{2}+1)$--fold degenerate if $n$ is even and $\frac12(n+1)$--fold degenerate if $n$ is odd \cite{Flu71}. We say that $\vert n, \ell \rangle_e$ and $\vert s,\ell\rangle$ respectively define the {\em energy space configuration} and the $(\ell,s)$-{\em configuration} of a given physical state $\vert \varphi \rangle$. In position-representation the wave-function $ \varphi_{n,\ell}(r,\theta) := \langle r,\theta \vert n,\ell \rangle_e
\equiv \langle r,\theta \vert s, \ell \rangle =\varphi_{s,\ell}(r,\theta)$ is determined by the product of the normalized functions (see Appendix~\ref{ApA} for details):
\bea
u_{s\ell} (r) := \langle r\vert s \rangle  =
\left[ \frac{2 \lambda^{\ell+3/2} \Gamma(s+1)}{\Gamma(s+\ell +\frac32)}\right]^{1/2} r^{\ell +1} e^{-\lambda r^2/2} L^{(\ell +1/2)}_s (\lambda r^2),
\nonumber\\[1ex]
\Theta_{\ell, m=0} (\theta) := \langle \theta \vert \ell \rangle = \left(
\frac{2\ell +1}{2}
\right)^{1/2}
P_{\ell} (\cos \theta),
\nonumber
\eea
where $L_n^{(\gamma)}(z)$ stands for the associated Laguerre polynomials and $P_{\ell}(z)$ for the Legendre Polynomials \cite{Olv10}. From now on we adopt the units for which $\lambda=1$. 

We look for a pair of first order operators $a_{\ell}$ and $a_{\ell}^{\dagger}$ such that
\be
H_{\ell} = a_{\ell}^{\dagger} a_{\ell} + \epsilon_{\ell}; \qquad a_{\ell}^{\dagger} := -\frac{d}{dr} + \alpha(r,\ell); \qquad (a_{\ell}^{\dagger})^{\dagger} =a_{\ell},
\label{f5}
\ee
with $\epsilon_{\ell}$ a real constant to be determined and $\alpha(r,\ell)$ a function satisfying the Riccati equation
\be
-\alpha' + \alpha^2 = V_{\ell} -\epsilon_{\ell}, \qquad f' = \frac{df}{dr}.
\label{f6}
\ee
Using $\alpha=-\frac{d}{dr} \ln v$, with $v(r,\ell)$ a function that is not necessarily of finite norm, the nonlinear equation (\ref{f6}) is transformed into the eigenvalue problem
\be
-v'' + \left[\frac{\ell(\ell+1)}{r^2} + r^2 -\epsilon_{\ell} \right] v=0.
\ee
The general solution of this last linear equation, with $2\mathtt{E}_n$ instead of $\epsilon_{\ell}$, is given in (\ref{ugral}) and serves as a repository seed of solutions to the nonlinear Riccati equation (\ref{f6}). As discussed below, there are two immediate particular solutions of (\ref{f6}) that give rise to four different factorizations.

\noindent
$\bullet$ {\bf First pair of intertwining relationships}

A particular solution of (\ref{f6}) is easily found by taking $\delta=0$ in (\ref{ugral}). Using this in (\ref{f5}) we have a pair of operators that do not commute
\be
a^{\dagger}_{\ell} a_{\ell}= H_{\ell} -\epsilon_{\ell}, \qquad a_{\ell} a^{\dagger}_{\ell}= H_{\ell +1} -\epsilon_{\ell -1}, \quad \epsilon_{\ell} = 2\ell +3.
\label{f10}
\ee

\begin{figure}[htb]
\centering 
\includegraphics[width=0.4\textwidth]{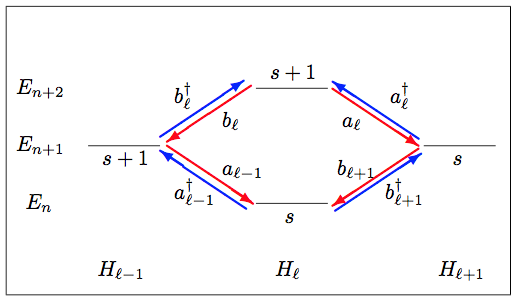} 

\caption{\footnotesize For fixed $\ell$, the eigenvectors of $H_{\ell}$ are intertwined with those of $H_{\ell\pm1}$ by the action of the factorizing operators.}
\label{enconf}
\end{figure}

\begin{figure}[htb]
\centering 
\includegraphics[width=0.35\textwidth]{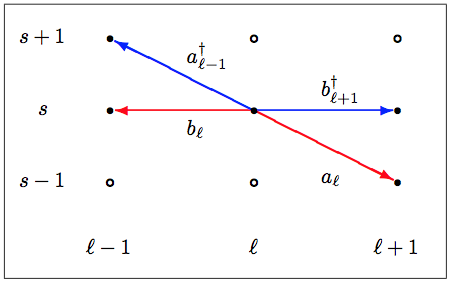}

\caption{\footnotesize Action of the factorizing operators in the $(\ell,s)$-plane. Note that $a_{\ell}$ and $a^{\dagger}_{\ell -1}$ operate in diagonal form by mapping $(\ell,s)$ into $(\ell+1,s-1)$ and $(\ell-1,s+1)$ respectively. In turn, $b_{\ell}$ and $b^{\dagger}_{\ell+1}$ operate in horizontal form by mapping $(\ell,s)$ into $(\ell-1,s)$ and $(\ell+1,s)$ respectively.
}
\label{slpic}
\end{figure}

Then, we have the intertwining relationships
\be
H_{\ell+1} a_{\ell} = a_{\ell} (H_{\ell} -2), \qquad  H_{\ell} a^{\dagger}_{\ell} = a^{\dagger}_{\ell} (H_{\ell +1} +2),
\nonumber
\ee
so that the eigenvectors of $H_{\ell}$ are intertwined with those of $H_{\ell \pm 1}$ by the action of $a_{\ell}$ and $a^{\dagger}_{\ell}$ (see Figure~\ref{enconf}). Accordingly,
\be
a^{\dagger}_{\ell-1} a_{\ell-1}= H_{\ell-1} -\epsilon_{\ell-1}, \qquad a_{\ell-1} a^{\dagger}_{\ell-1}= H_{\ell} -\epsilon_{\ell -2}.
\label{lost1}
\ee
The action of $a_{\ell}$ and $a_{\ell-1}^{\dagger}$ on the eigenvectors of $H_{\ell}$ is as follows\footnote{The vector $a^{\dagger}_{-1} \vert s, 0 \rangle$ is {\em unphysical} because it does not satisfy the boundary conditions in position-representation.}:
\be
\begin{array}{c}
a_{\ell} \vert s, \ell \rangle = 2\sqrt{s} \vert s-1, \ell+1 \rangle, \qquad a_{\ell} \vert 0, \ell \rangle =  0,\\[2ex]
a_{\ell-1}^{\dagger} \vert s, \ell \rangle = 2\sqrt{s+1} \vert s+1, \ell-1 \rangle, \qquad  a^{\dagger}_{-1} \vert s, 0 \rangle = \mbox{unphysical}.
\end{array}
\ee
Thus, the action of $a_{\ell}$ $(a^{\dagger}_{\ell})$ decreases (increases) the radial quantum number $s$ in one unit at the price of increasing (decreasing) the orbital quantum number $\ell$ in one unit. Considering the $(\ell,s)$-plane integrated by the quantum numbers $\ell$ and $s$, these last operations can be represented by either the diagram shown in Figure~\ref{slpic} or the mappings
\be
a_{\ell}: (\ell,s) \mapsto (\ell+1,s-1), \qquad a^{\dagger}_{\ell-1}: (\ell,s) \mapsto (\ell-1,s+1).
\nonumber
\ee

\noindent
These last results motivate the introduction of a pair of free-index operators 
\be
\begin{array}{l}
a^+ \vert s, \ell \rangle =\tfrac12 a_{\ell-1}^{\dagger} \vert s, \ell \rangle  =\sqrt{s+1} \,  \vert s+1, \ell -1 \rangle,\\[2ex]
a^- \vert s, \ell \rangle = \tfrac12 a_{\ell} \vert s, \ell \rangle =\sqrt{s}\,  \vert s-1, \ell +1 \rangle,
\end{array}
\label{freea}
\ee
so that
\be
a^+ a^- \vert s, \ell \rangle = s \vert s, \ell \rangle, \quad a^- a^+ \vert s, \ell \rangle = (s+1) \vert s, \ell \rangle \quad \Rightarrow \quad N_s \vert s, \ell \rangle = s \vert s, \ell \rangle,
\nonumber
\ee
with $N_s:=a^+ a^-$ the (principal quantum) number operator. In this form we arrive at the boson algebra
\be
[a^-, a^+ ]= \mathbb I, \quad [N_s, a^{\pm} ]= \pm a^{\pm},
\label{bosalg1}
\ee
where $\mathbb I$ stands for the identity operator.

\noindent
$\bullet$ {\bf Second pair of intertwining relationships}

Another particular solution of (\ref{f6}) is obtained if $\gamma=0$ in (\ref{ugral}). The resulting $\alpha$-function leads to a new pair of factorizing operators that do not commute
\be
b^{\dagger}_{\ell} b_{\ell}= H_{\ell} +\epsilon_{\ell-2}, \qquad b_{\ell} b^{\dagger}_{\ell}= H_{\ell -1} +\epsilon_{\ell -1}.
\label{f25}
\ee
Thus, we obtain the intertwining relationships
\be
H_{\ell -1} b_{\ell} = b_{\ell} (H_{\ell} -2), \qquad H_{\ell} b^{\dagger}_{\ell} = b^{\dagger}_{\ell} (H_{\ell-1}+2).
\label{f25a}
\ee
As in the previous case, it is convenient to rewrite (\ref{f25}) and (\ref{f25a}) as follows
\bea
b^{\dagger}_{\ell+1} b_{\ell+1}= H_{\ell+1} +\epsilon_{\ell-1}, \qquad b_{\ell+1} b^{\dagger}_{\ell+1}= H_{\ell } +\epsilon_{\ell },
\label{lost2}\\[1ex]
H_{\ell -1} b_{\ell} = b_{\ell} (H_{\ell}-2), \quad H_{\ell+1} b^{\dagger}_{\ell+1} = b^{\dagger}_{\ell+1} (H_{\ell}+2).
\eea
In this form (see Figures~\ref{enconf} and \ref{slpic}),
\be
b_{\ell}: (\ell,s) \mapsto (\ell-1,s), \qquad b^{\dagger}_{\ell+1}: (\ell,s) \mapsto (\ell+1,s),
\nonumber
\ee
and
\be
\begin{array}{c}
b_{\ell} \vert s, \ell \rangle = 2\sqrt{s+\ell +\tfrac12} \vert s, \ell -1 \rangle, \qquad b_0 \vert s, 0 \rangle= \mbox{unphysical},\\[3ex]
b_{\ell+1}^{\dagger} \vert s, \ell \rangle = 2\sqrt{s+\ell +\tfrac32} \vert s, \ell +1 \rangle.
\end{array}
\ee
The iterated action of $b_{\ell}$ and $b^{\dagger}_{\ell}$ can be managed in terms of the free-index operators
\be
\begin{array}{l}
b^+ \vert s, \ell \rangle = \tfrac12 b^{\dagger}_{\ell+1} \vert s, \ell \rangle = \sqrt{\ell + s+ \tfrac32} \, \vert s, \ell +1\rangle,\\[2ex]
b^- \vert s, \ell \rangle = \tfrac12 b_{\ell} \vert s, \ell \rangle = \sqrt{\ell + s+ \tfrac12} \, \vert s, \ell -1 \rangle,\\[2ex]
N_{\ell}  \vert s, \ell \rangle  =  (\ell + s+ \tfrac12) \vert s, \ell \rangle,
\end{array}
\label{freeb}
\ee
with $N_{\ell} = b^+ b^-$ the (orbital quantum) number operator. These last satisfy the boson algebra
\be
[b^-, b^+]= \mathbb I, \quad [N_{\ell}, b^{\pm} ] = \pm b^{\pm}.
\label{bosalg2}
\ee

\noindent
$\bullet$ {\bf Canonical factorizations and hierarchies of states}

The factorizations (\ref{f10}), (\ref{lost1}), (\ref{f25}) and (\ref{lost2}) were already noticed in \cite{Fer96} (see also \cite{Cab08}). Other factorizations, named after Mielnik \cite{Mie84}, are easily obtained by taking $\alpha =-\frac{d}{dr} \ln v$, with $\gamma$ and $\delta$ properly chosen in the general function (\ref{ugral}). At this stage, we would like to emphasize that our approach can be easily extended to a supersymmetric model of the radial oscillator because, as it is well known, the factorization method is in the kernel of supersymmetric quantum mechanics (see, e.g., the review papers \cite{Mie04,And12,Kha14,Suk14,Fer14}). 

For purposes that will be clear in the sequel, it is convenient to classify the states of the radial oscillator as follows:

\begin{enumerate}
\item
{\em Definite orbital angular momentum states.} For fixed $\ell$ we have an infinite number of possible energy eigenvalues and eigenvectors, 
\be
\vert s, \ell \rangle \equiv \vert s\rangle_{\ell}, \quad E_{s,\ell} = \epsilon_{\ell} +4s, \quad s=0,1,\ldots
\nonumber
\ee
We say that the infinite-dimensional space ${\cal H}_{\ell}$ spanned by the vectors $\vert s \rangle_{\ell}$ is an $\ell$-hierarchy of definite orbital angular momentum states.

\item
{\em Definite energy states.} For fixed $n$ one has a finite number of possible energy eigenvalues and eigenvectors
\be
\vert n,\ell \rangle_e, \quad E_{n} = 2n +3, \quad n=2s+\ell =0,1,\ldots
\nonumber
\ee
We say that the finite-dimensional space ${\cal H}_{(n)}$ spanned by the vectors $\vert n, \ell \rangle_e$ is an $E_{n}$-hierarchy of definite energy states. The dimension of ${\cal H}_{(n)}$ is equal to the accidental degeneracy $d= \mbox{deg}({E}_n)$ of the related energy eigenvalue.
 
\end{enumerate}

\section{Intertwining algebras and coherent states}
\label{intertwining}

The intertwining relationships defined in Section~\ref{secfac} connect the elements of a given $\ell$-hierarchy with those of the $\ell \pm 1$ hierarchies, and the elements of the $E_n$-hierarchy with the ones of the $E_{n\pm1}$-hierarchies. In this section we look for relationships intertwining the elements of a given hierarchy with other elements of the same hierarchy. The main idea is to find the combinations of operators $a^{\pm}$ and $b^{\pm}$ that produce vertical, horizontal or diagonal mappings $s \leftrightarrow s+1$ in  the $(\ell, s)$-plane.

\subsection{Spectrum generating algebra of definite orbital angular momentum hierarchies}
\label{algdm}

\begin{figure}[htb]
\centering 
\includegraphics[width=0.35\textwidth]{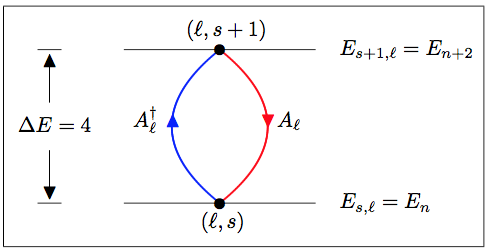}

\caption{\footnotesize Action of the definite angular momentum ladder operators $A_{\ell}$ and $A^{\dagger}_{\ell}$. In the energy space configuration they preserve the angular momentum but decrease and increase the energy eigenvalue in four units respectively. In the $(\ell,s)$-plane, the action of $A_{\ell}$ ($A^{\dagger}_{\ell}$ ) annihilates (creates) a node in the position representation of the definite angular momentum vector $\vert s\rangle_{\ell}$.
}
\label{Aops}
\end{figure}

Using the (inner) red and (outer) blue arrows in Figure~\ref{enconf} we get the operators
\be
A_{\ell} =a_{\ell -1}b_{\ell} = b_{\ell+1}a_{\ell} , \qquad A^{\dagger}_{\ell} =b^{\dagger}_{\ell} a^{\dagger}_{\ell -1} = a^{\dagger}_{\ell} b^{\dagger}_{\ell+1} .
\nonumber
\ee
As the action of  $b_0$ and $a^{\dagger}_{-1}$ on the states $\vert s, 0 \rangle$ gives unphysical vectors, we shall use the expressions
\be
A_{\ell} = b_{\ell+1}a_{\ell} , \qquad A^{\dagger}_{\ell} = a^{\dagger}_{\ell} b^{\dagger}_{\ell+1} , \quad (A^{\dagger}_{\ell})^{\dagger} = A_{\ell}.
\label{ladA}
\ee
The action of these operators is depicted in Figure~\ref{Aops}, they decrease or increase the radial quantum number $s$ in one unit:
\be
A_{\ell}:  (\ell,s) \mapsto (\ell, s-1), \qquad A^{\dagger}_{\ell}:  (\ell,s) \mapsto (\ell, s+1).
\nonumber
\ee

\noindent
Together with $H_{\ell}$, these ladder operators satisfy the commutation rules
\be
[H_{\ell}, A_{\ell}]=-4A_{\ell}, \quad [H_{\ell}, A^{\dagger}_{\ell}]=4A^{\dagger}_{\ell}, \quad [A_{\ell}, A^{\dagger}_{\ell}]=8H_{\ell}.
\label{alg1}
\ee
Hence, considering the free-index operators
\be
L_+ = a^+ b^+, \quad L_- =b^- a^-, \quad L_3 =\tfrac12 (N_s + N_{\ell} +1)
\label{S1a}
\ee
and the identification
\be
L_+ \leftrightarrow \tfrac14 A_{\ell}^{\dagger}, \quad L_-  \leftrightarrow \tfrac14 A_{\ell}, \quad L_3 \leftrightarrow \tfrac14 H_{\ell},
\label{S1b}
\ee
the commutators (\ref{alg1}) correspond to the Lie algebra of $SU(1,1)$,
\be
[L_3, L_{\pm}]=\pm L_{\pm}, \quad [L_-,L_+]=2L_3.
\label{S1c}
\ee
Remarkably, we can write
\be
L_3 = r L_3^M, \quad L_{\pm}= -r L_{\pm}^M,
\nonumber
\ee
with $L_3^M$ and $L_{\pm}^M$ the $su(1,1)$ generators reported by Quesne and Moshinsky for a 3D harmonic oscillator with $\ell$ fixed (after integrating out the angular variables) \cite{Mos71}. The factor $r$ at the left of the Moshinsky operators obeys the fact that the treatment in \cite{Mos71} is based on the radial functions $R_{n\ell}(r)$ while ours is developed on the functions  $u_{n\ell} (r) =rR_{n\ell} (r)$, see Appendix~\ref{ApA}. The above expressions show that $SU(1,1)$ is the generating group of the radial oscillator. Considering the Moshinsky Casimir of the algebra $C^M = \tfrac14 ({\mathbf L}^2 - \tfrac34)$, with $\mathbf L^2$ the square of the orbital angular momentum operator, we see that the Bargmann index $\kappa$ of the representation is defined by the factorization constant $\kappa= \tfrac14 (2\ell +3) = \tfrac14 \epsilon_{\ell}$. That is, for the two-mode boson realization (\ref{S1a}), we have $\kappa= \tfrac34, \tfrac54, \tfrac94, \ldots$ 

The space of states of the radial oscillator decomposes into the direct sum $\bigoplus_{\kappa=3/4}^{\infty} {\cal H}^{\kappa}$, where the `vertical' subspaces ${\cal H}^{\kappa}$ are the $\ell$-hierarchies of definite angular momentum ${\cal H}_{\ell}$ introduced in Section~\ref{secfac}. 

On the other hand, expressions (\ref{alg1})--(\ref{S1c}) are the Schwinger representation of the $su(1,1)$ Lie algebra for which the definite angular momentum operators $L_{\pm}$ are linked to the two bosons $a^{\pm}$ and $b^{\pm}$, whenever $N_s + N_{\ell} + 1=2L_3$, see \cite{Sch52}. In this context, the boson occupation $2L_3$ gives the number $2(s + \kappa)$. Thus, given $\kappa$, the representation is labelled by the radial quantum number $s$. In addition, the action of $L_{\pm}$ on the vectors belonging to the $\ell$-hierarchy is given by
\be
L_- \vert s \rangle_{\ell} = \sqrt{s(s+\ell+1/2)} \vert s-1 \rangle_{\ell}, \quad L_+ \vert s\rangle_{\ell} = \sqrt{(s+1)(s+\ell+3/2)} \vert s+1 \rangle_{\ell}.
\label{eles}
\ee
That is, acting on the points in the $(\ell,s)$-plane, the operator $L_-$ ($L_+$) produces vertical displacements by decreasing (increasing) the radial quantum number $s$ in one unit, as desired. Moreover, $s \mapsto s\pm 1$ leads to $n \mapsto n\pm 2$, so that the action of  $A_{\ell}$ and $A^{\dagger}_{\ell}$ depicted in Figure~\ref{Aops}  produces jumps of energy in steps of four units just because there are two photons  involved in the energy transitions (remember that $E_n$ is twice the dimensionless energy $\mathtt{E}_n$ of the spherical oscillator).

For completeness, besides the commutation rules (\ref{bosalg1}) and (\ref{bosalg2}), in the Schwinger representation we have
\be
\begin{array}{c}
[a^-, b^-] = [a^+, b^+]=[a^-,b^+]=[b^-,a^+]=0, \quad [L_{\pm}, a^{\pm}] = [L_{\pm},b_{\pm}]=0,\\[1ex]
[L_{\pm}, a^{\mp}] =\mp b^{\pm}, \quad [L_{\pm}, b^{\mp} ]= \mp a^{\pm}, \quad [L_3, a^{\pm}]= \pm \tfrac12 a^{\pm}, \quad [L_3, b^{\pm}]= \pm \tfrac12 b^{\pm}.
\end{array}
\label{S1d}
\ee

\subsubsection{$SU(1,1)$ Barut-Girardelo coherent states}
\label{sec311}

The Barut-Girardelo \cite{Bar71} eigenvalue equation
\be
L_-\vert z \rangle_{BG} = z \vert z \rangle_{BG}, \quad z\in \mathbb C,
\nonumber
\ee
can be solved by using (\ref{eles}) and the appropriate combination of definite angular momentum states $\vert s \rangle_{\ell}$. We obtain
\be
\vert z \rangle_{BG} =\frac{ \vert z \vert^{\frac{2\ell +1}{4}}}{\sqrt{I_{\ell +\frac12}(2\vert z \vert)}} \sum_{s=0}^{\infty} \frac{z^s}{\sqrt{\Gamma(s+1) \Gamma(s+\ell+3/2)}} \vert s \rangle_{\ell}
\label{bg2}
\ee
with $I_{\nu}(z)$ the modified Bessel function of the first kind \cite{Olv10}, Eq.~10.25.2,
\be
I_{\nu} (z) = \left(\tfrac12 z \right)^{\nu} \sum_{k=0}^{\infty} \frac{\left( \tfrac14 z^2\right)^k}{\Gamma(k+1) \Gamma(k+1+\nu)}.
\nonumber
\ee
In position representation the ket (\ref{bg2}) acquires the form
\be
 \varphi_z^{BG}(r):= \langle r \vert z \rangle_{BG}= \left[\frac{ 2}{I_{\ell +\frac12}(2\vert z \vert)} \right]^{1/2}  \vert z \vert^{\frac{2\ell +1}{4}}  r^{\ell +1} e^{-r^2/2} \sum_{s=0}^{\infty} \frac{z^s}{\Gamma(s+\ell +3/2)} L_s^{\ell +\frac12} (r^2).
\label{bg3}
\ee
Similar results have been reported for the Calogero-Sutherland oscillator in \cite{Aga95} (see also \cite{Sch16}).  We can further simplify our expressions by using the Eq.~8.975.3 of Ref.~\cite{Gra07},
\be
\frac{e^z}{(xz)^{\nu/2}} J_{\nu} (2 \sqrt{xz})= \sum_{k=0}^{\infty} \frac{z^k}{\Gamma(k+\nu+1)} L_k^{\nu} (x),
\nonumber
\ee
to perform the sum in (\ref{bg3}). Up to a global phase it gives
\be
\varphi_z^{BG}(r)=\left[\frac{2 r}{I_{\ell +\frac12}(2\vert z \vert)} \right]^{1/2} e^{z -r^2/2} J_{\ell+1/2}(2r \sqrt{z}),
\label{bg4}
\ee
with $J_{\nu}(z)$ the Bessel function of the first kind \cite{Olv10}. 

\begin{figure}[htb]
\centering 
\begin{subfigure}[b]{.2\linewidth}
\centering
\includegraphics[width=0.99\textwidth]{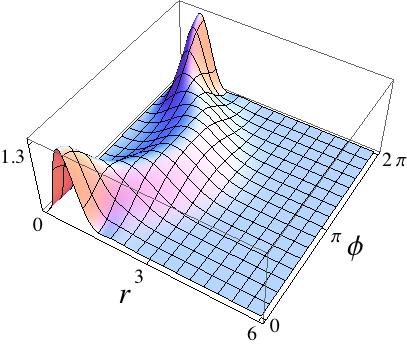} 
\caption{$\ell=0, \vert z \vert =1$}
\end{subfigure}
\begin{subfigure}[b]{.2\linewidth}
\includegraphics[width=0.99\textwidth]{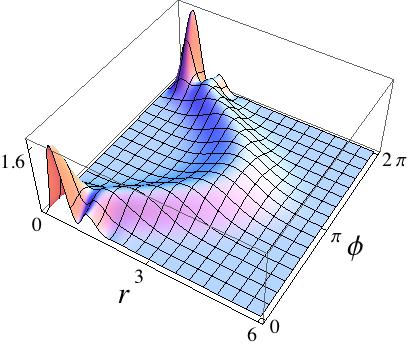} 
\caption{$\ell=0, \vert z \vert =3$}
\end{subfigure}
\begin{subfigure}[b]{.2\linewidth}
\includegraphics[width=0.99\textwidth]{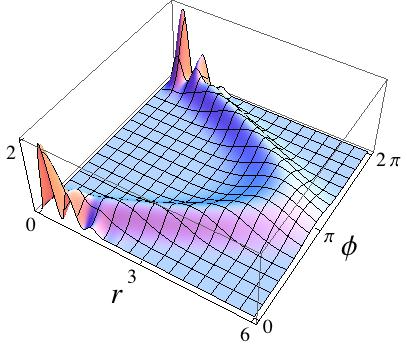}
\caption{$\ell=0, \vert z \vert =6$}
\end{subfigure}
\vskip2ex
\begin{subfigure}[b]{.2\linewidth}
\centering
\includegraphics[width=0.99\textwidth]{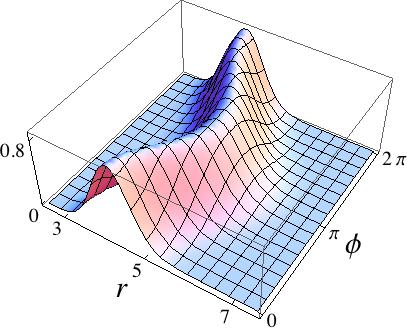} 
\caption{$\ell=20, \vert z \vert =2$}
\end{subfigure}
\begin{subfigure}[b]{.2\linewidth}
\includegraphics[width=0.99\textwidth]{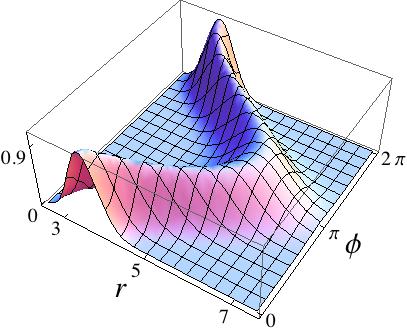} 
\caption{$\ell=20, \vert z \vert =6$}
\end{subfigure}
\begin{subfigure}[b]{.2\linewidth}
\includegraphics[width=0.99\textwidth]{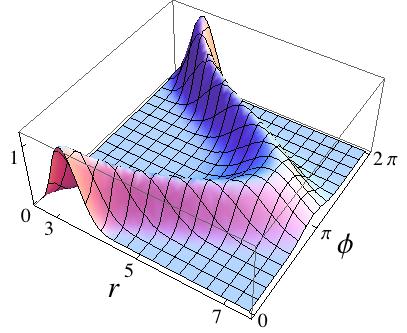}
\caption{$\ell=20, \vert z \vert =8$}
\end{subfigure}

\caption{\footnotesize Probability density of the Barut-Girardello coherent states $\varphi_z^{BG}(r)$ defined in (\ref{bg4}) for the indicated values of $z=\vert z \vert e^{-i\phi}$. The orbital quantum number $\ell$ has been fixed as $\ell=0$ (first row) and $\ell=20$ (second row).}
\label{BG1}
\end{figure}

The global behaviour of the probability density $\vert \varphi_z^{BG}(r) \vert^2$ is depicted in Figure~\ref{BG1} for the $\ell$-hierarchies defined by $\ell =0$, $\ell=20$, and the indicated values of the complex eigenvalue $z=\vert z \vert e^{-i\phi}$. For $\ell$ and $\vert z \vert$ fixed, the closer the phase $\phi$ is to either 0 or $2\pi$, the nearer to the origin is the center of the probability density. The largest distance between the origin and the center of the probability density is reached for $\phi=\pi$ and increases as $\vert z \vert$. On the other hand, given $\ell$ the probability density exhibits more than one peak for $\vert z \vert >> \ell$, as it is shown in Figures~\ref{BG1}(a--c).

An additional form of $\vert z \rangle_{BG}$ is available by writing the definite angular momentum vector $\vert s \rangle_{\ell}$ as the result of the iterated action of $L_+$ on the ground state $\vert 0 \rangle_{\ell}$, that is
\be
\vert s \rangle_{\ell} = \left[ \frac{\Gamma(\ell+3/2)}{\Gamma(s+1) \Gamma(s+\ell+3/2)} \right]^{1/2} L^s_+ \vert 0 \rangle_{\ell}.
\label{bggral}
\ee
The introduction of this last expression into (\ref{bg2}) produces
\be
\vert z \rangle_{BG} = \vert z \vert^{\frac{2\ell +1}{4}}
\left[ \frac{\Gamma(\ell+3/2)}{I_{\ell +\frac12}(2\vert z \vert)} \right]^{1/2}
\hat I (zL_+) \vert 0 \rangle_{\ell},
\nonumber
\ee
with $\hat I (zL_+)$ the Bessel-like operator
\be
\hat I (zL_+):= \sum_{s=0}^{\infty} \frac{(zL_+)^s}{\Gamma(s+1) \Gamma(s+\ell+3/2)} \circeq  (zL_+)^{-\frac{2\ell +1}{4}} I_{\ell +\frac12} (2\sqrt{zL_+}).
\nonumber
\ee

\begin{figure}[htb]
\centering 
\includegraphics[width=0.3\textwidth]{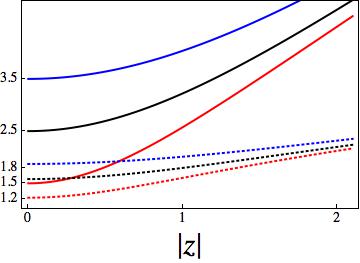} 

\caption{\footnotesize Minimum uncertainty $ \frac12 \langle L_3 \rangle_{BG} = \Delta L_1 \Delta L_2$ satisfied by the Barut-Girardelo coherent states $\vert z \rangle_{BG}$ for $\ell=0$ (red, bottom), $\ell=1$ (black) and $\ell=2$ (blue, top). The dotted curves correspond to the deviations $\Delta L_1 = \Delta L_2$. Note that in all cases the variances $(\Delta L_{1,2})^2$ evaluated in $\ell \neq 0$ are bigger than the one in $\ell=0$.
}
\label{BG3}
\end{figure}

\noindent
$\bullet$ {\bf Uncertainty relation.} It is well known that $L_3$ and  the ladder operators $L_{\pm}$ give rise to the inequality
\be
\Delta L_1 \Delta L_2 \geq \frac12 \langle L_3 \rangle
\label{uncer1}
\ee
that is fulfilled by the quadratures
\be
L_1 =\frac12 (L_+  + L_-), \quad L_2 =\frac{1}{2i} (L_+-L_-).
\label{mosh1}
\ee
In our case these last operators have the following position-representation 
\be
L_1 = -\frac14 H_\ell + \frac12 r^2, \quad L_2 = - \frac i4 \left(2r\frac{d}{dr} + 1\right),
\label{mosh2}
\ee
which is consistent with the realization of the $su(1,1)$ Lie algebra reported in \cite{Mos71}. Using the coherent states (\ref{bg2}) we get
\be
(\Delta L_1)^2 = (\Delta L_2)^2 = \ell + \frac32 + 2 \vert z \vert \frac{I_{\ell + 3/2}(2 \vert z \vert)}{I_{\ell + 1/2}(2 \vert z \vert)} = \frac12 \langle L_3 \rangle_{BG} = \Delta L_1 \Delta L_2.
\label{min1}
\ee
That is, $\vert z \rangle_{BG}$ are minimum uncertainty states (see Figure~\ref{BG3})  of average energy $ \overline E_{\ell} := \langle H_{\ell} \rangle_{BG} = 4 \langle L_3 \rangle_{BG}$ defined by $\vert z \vert$, as indicated in Eq.~(\ref{min1}). 

\begin{figure}[htb]
\centering 
\begin{subfigure}[b]{.2\linewidth}
\centering
\includegraphics[width=0.99\textwidth]{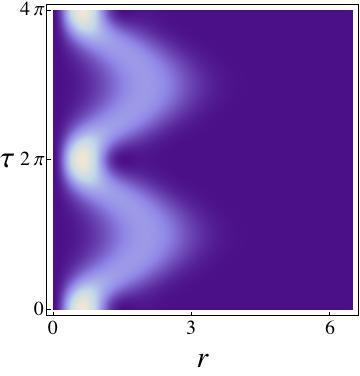} 
\caption{$\ell=0, \vert z \vert =1$}
\end{subfigure}
\begin{subfigure}[b]{.2\linewidth}
\includegraphics[width=0.99\textwidth]{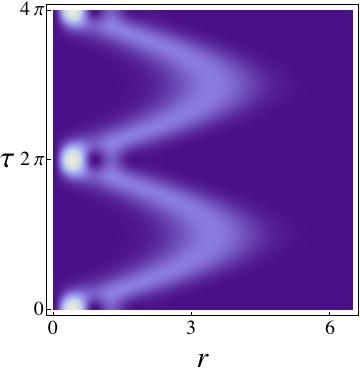} 
\caption{$\ell=0, \vert z \vert =3$}
\end{subfigure}
\begin{subfigure}[b]{.2\linewidth}
\includegraphics[width=0.99\textwidth]{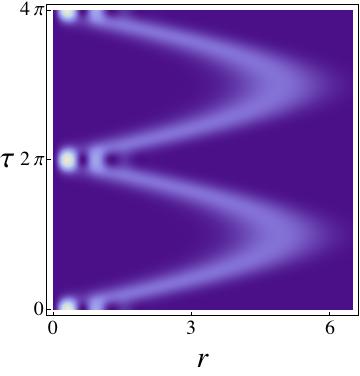}
\caption{$\ell=0, \vert z \vert =6$}
\end{subfigure}
\vskip2ex
\begin{subfigure}[b]{.2\linewidth}
\centering
\includegraphics[width=0.99\textwidth]{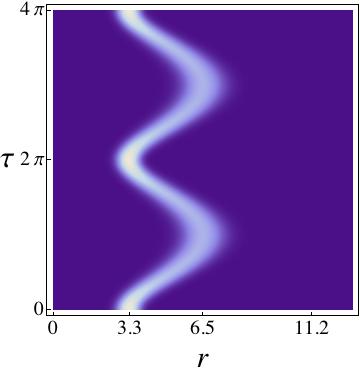} 
\caption{$\ell=20, \vert z \vert =8$}
\end{subfigure}
\begin{subfigure}[b]{.2\linewidth}
\includegraphics[width=0.99\textwidth]{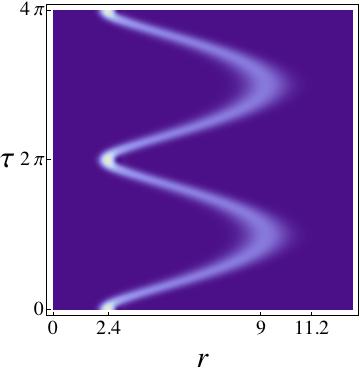} 
\caption{$\ell=20, \vert z \vert =20$}
\end{subfigure}
\begin{subfigure}[b]{.2\linewidth}
\includegraphics[width=0.99\textwidth]{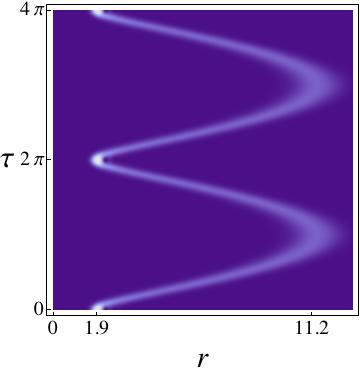}
\caption{$\ell=20, \vert z \vert =32$}
\end{subfigure}

\caption{\footnotesize Time-evolution of the probability density $\vert \varphi_z^{BG}(r,t) \vert^2$ for $\ell=0$ (first row), $\ell=20$ (second row), and the indicated values of $\vert z \vert$. In all cases $\phi=0$ is the initial condition and the vertical axis measures time in units of $\tau=2\lambda t$. The wave packet propagates in time by oscillating back and forth between two turning points that depend on $\vert z \vert$. At $\tau =2 k \pi$, $k \in \mathbb Z^+$, for $\vert z \vert >>\ell$, the probability density exhibits a series of peaks that are disseminated close to the origin, see figures (b--c) and (f). The packet becomes broad and single peaked as its center approaches the outer turning point. }
\label{BG2}
\end{figure}

\vskip1ex
\noindent
$\bullet$ {\bf Time evolution and classical behaviour.} The time-evolution of the coherent state (\ref{bg2}) is easily shown to be
\be
\vert z; t \rangle_{BG} = e^{-iH_{\ell}t} \vert z\rangle_{BG}= e^{-i\lambda(2\ell +3)t} \vert z(t) \rangle_{BG}, \quad z(t) =ze^{-i4\lambda t},
\nonumber
\ee
where the parameter $\lambda = \tfrac{m\omega}{\hbar}$ has been retrieved for clarity. That is, the time-dependence of our coherent state is encoded in the phase of the complex eigenvalue $z(t)=\vert z \vert e^{-i(\phi + 4\lambda t)}$ with $z(0)=z$, as usual. With this result in (\ref{bg4})  we see that the time-evolution of the probability density $\vert \varphi_z^{BG}(r,t) \vert^2$ is periodic with angular frequency $2\lambda$ (twice the oscillation frequency $\omega$). The center of the wave-packet oscillates back and forth between two {\em turning points} that depend on $\vert z \vert$ (see Figure~\ref{BG2}) and can be determined by matching the average energy $\overline E^{BG}_{\ell}$ with the effective potential:
\be
\overline E^{BG}_{\ell} = \frac{\ell (\ell+1)}{r^2} + r^2.
\nonumber
\ee

\noindent
Moreover, by construction $\langle L_- \rangle_{BG}=z$ and $\langle L_+ \rangle_{BG}= z^*$, with $z^*$ the complex conjugate of $z \in \mathbb C$, so we immediately have $\langle L_1 \rangle_{BG} = \mbox{Re}(z) = \vert z \vert \cos \phi$ and $\langle L_1 \rangle_{BG} = \mbox{Im}(z) = \vert z \vert \sin \phi$. Considering the time-dependence $\phi \rightarrow \phi + 4\lambda t$ we see that $\langle L_1 \rangle_{BG}$ and $\langle L_2 \rangle_{BG}$ behave as the classical generalized coordinates associated with a harmonic oscillator of period $T=\tfrac{\pi}{4\lambda}$. An interpretation of these results is as follows. The center of the packet $\varphi^{BG}_z(r,t)$ behaves as a classical particle in the radial oscillator potential and describes elliptical paths by oscillating twice between the turning points during one complete period.

\begin{figure}[htb]
\centering 
\begin{subfigure}[b]{.3\linewidth}
\centering
\includegraphics[width=0.9\textwidth]{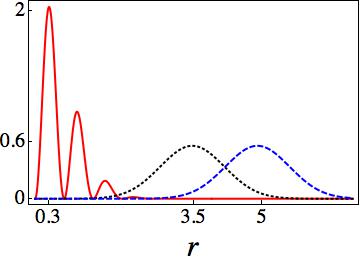} 
\caption{$\ell=0, \vert z \vert =6$}
\end{subfigure}
\hskip3ex
\begin{subfigure}[b]{.3\linewidth}
\centering
\includegraphics[width=0.9\textwidth]{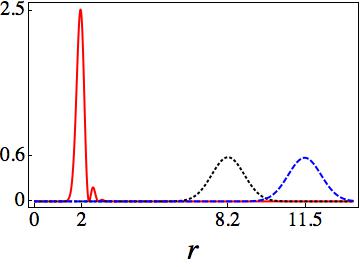}
\caption{$\ell=20, \vert z \vert =32$}
\end{subfigure}

\caption{\footnotesize For $\vert z \vert >> \ell$, the number of peaks of the probability density $\vert \varphi_z^{BG}(r,t) \vert^2$ depends on the phase of the time-evolved complex eigenvalue $z(\tau) = \vert z \vert e^{-i\tau}$, see Figure~\ref{BG2}. At $\tau =0$ (red continuous curve), the peaks are disseminated in the vicinity of the origin. The probability density becomes single peaked and broader as its center approaches the outer turning point; compare the initial density profile with the ones at $\tau = \frac{\pi}{2}$ (dotted black curve) and $\tau = \pi$ (dashed blue curve).  
}
\label{BG2A}
\end{figure}

On the other hand, the probability density becomes narrow as it approaches the origin and spreads about the outer turning point (see Figure~\ref{BG2}). The phenomenon may be interpreted as a generalized squeezing due to the centrifugal barrier of the potential. If $\vert z \vert >> \ell$, the probability density exhibits a series of local maxima (peaks), the number of which depends on the value of the phase $\phi$ and evolves in time periodically, as it is shown in Figures~\ref{BG2} and \ref{BG2A}.

\subsubsection{$SU(1,1)$ Perelomov coherent states}
\label{su11per}

\begin{figure}[htb]
\centering 
\begin{subfigure}[b]{.2\linewidth}
\centering
\includegraphics[width=0.99\textwidth]{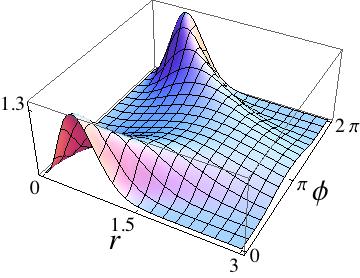} 
\caption{$\ell=0$}
\end{subfigure}
\begin{subfigure}[b]{.2\linewidth}
\includegraphics[width=0.99\textwidth]{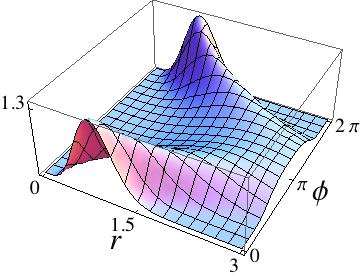} 
\caption{$\ell=1$}
\end{subfigure}
\begin{subfigure}[b]{.2\linewidth}
\includegraphics[width=0.99\textwidth]{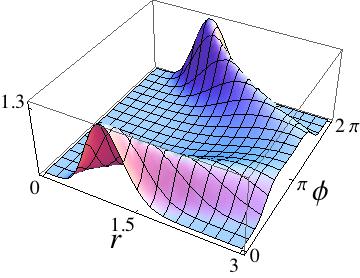}
\caption{$\ell=2$}
\end{subfigure}
\caption{\footnotesize Probability density of the Perelomov coherent state $\varphi_z^{P}(r)$ defined in (\ref{mos1}) for the indicated $\ell$-hierarchies. In all cases $z= \vert z \vert e^{-i\phi}$ is such that $\vert z \vert=0.5$ and $\phi \in [0,2\pi)$.
}
\label{P1}
\end{figure}

Following the Perelomov approach \cite{Per86}, the $\ell$-hierarchy coherent states are of the form
\be
\vert z \rangle_P = (1-\vert z \vert^2)^{\frac{2\ell+3}{4}} \sum_{s=0}^{\infty} \left[ \frac{\Gamma(s+\ell+3/2)}{\Gamma(s+1) \Gamma(\ell+3/2)} \right]^{1/2} z^s \vert s \rangle_{\ell}.
\ee
Below we show that $\vert z \rangle_P$ are indeed displaced versions of the appropriate {\em extremal state}. These vectors have the  position-representation
\be
\varphi^P_z(r) = \left[ \frac{2}{\Gamma(\ell+3/2)} \right]^{1/2}
(1-\vert z \vert^2)^{\frac{2\ell+3}{4}}
r^{\ell+1} e^{-r^2/2} \sum_{s=0}^{\infty} z^s L_s^{\ell +\frac12} (r^2).
\nonumber
\ee
The above series converges in the unit disk $\vert z \vert <1$. Indeed, using the Eq.~8.975.1 of Ref.~\cite{Gra07},
\be
\frac{e^{\frac{xz}{z-1}}}{(1-z)^{\nu +1}} = \sum_{k=0}^{\infty} z^k L_k^{\nu} (x), 
\nonumber
\ee
one gets
\be
\varphi^P_z(r) = \left[ \frac{2}{\Gamma(\ell+3/2)} \right]^{1/2}  \left[ \frac{\sqrt{1-\vert z \vert^2}}{1-z}
\right]^{\frac{2\ell+3}{2}} r^{\ell+1} \exp\left[ -\frac{r^2}{2} \left(\frac{1+z}{1-z}\right)\right].
\label{mos1}
\ee
Clearly, these coherent states are single peaked for any value of $\ell$ and $\phi$ (see Figure~\ref{P1} and compare with Figure~\ref{BG1}). 

\noindent
The expression (\ref{mos1}) has been linked with the so called `radial coherent states' reported in \cite{Ger88}. There,  the authors use the Moshinky realization of the $su(1,1)$ algebra for the isotropic oscillator \cite{Mos71} as point of departure. Here, such a realization corresponds to the position-representation (\ref{mosh2}) of the quadratures (\ref{mosh1}) associated with the definite angular momentum operators $L_{\pm}$, as this has been discussed in the previous section.

To verify that $\vert z \rangle_P$ is the result of a displacement let us rewrite it in the form
\be
\vert z \rangle_P = (1-\vert z \vert^2)^{\frac{2\ell+3}{4}} e^{zL_+} \vert 0 \rangle_{\ell} =(1-\vert z \vert^2)^{\frac{2\ell+3}{4}} e^{zL_+} e^{-\overline z L_-} \vert 0 \rangle_{\ell},
\ee
where we have used (\ref{bggral}). Then, applying the well known disentangling formula \cite{Per86},
\be
e^{\xi L_+ -\overline \xi L_-} =e^{zL_+} e^{\ln (1-\vert z \vert^2) L_3} e^{-\overline z L_-}, \quad z=\tfrac{\xi}{\vert \xi \vert} \tanh (\vert \xi \vert), \quad \xi \in \mathbb C,
\label{etiq}
\ee
one immediately gets
\be
\vert z\rangle_P = D(\xi) \vert 0 \rangle_{\ell} \quad \mbox{with} \quad D(\xi)=e^{\xi L_+ -\overline \xi L_-}.
\label{operator}
\ee
That is, the Perelomov coherent states $\vert z \rangle_P$ are displaced versions of the `vacuum' definite angular momentum state $\vert 0 \rangle_{\ell}$ (remember, the radial quantum number $s=0$ means zero nodes in position-representation). This is precisely the reason for which the related probability density is single peaked for any value of the orbital quantum number $\ell$ and the phase $\phi$. A fact unnoticed in the literature on the matter already published (see e.g., \cite{Ger88}).

\begin{figure}[htb]

\centering 
\begin{subfigure}[b]{.2\linewidth}
\centering
\includegraphics[width=0.9\textwidth]{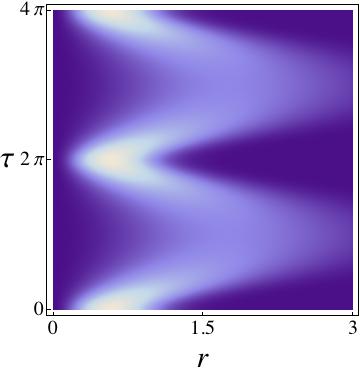} 
\caption{$\tau\in [0,4\pi]$}
\end{subfigure}
\hskip3ex
\begin{subfigure}[b]{.3\linewidth}
\centering
\includegraphics[width=0.9\textwidth]{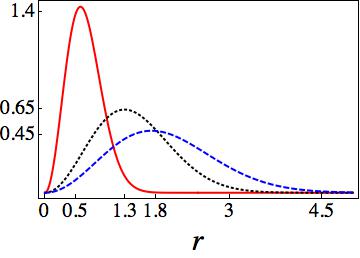}
\caption{$\tau=0,\tfrac{\pi}{2}, \pi$}
\end{subfigure}

\caption{\footnotesize Time-evolution of the probability density $\vert \varphi_z^{P}(r,t) \vert^2$ for  $\ell=0$, $\vert z \vert=0.5$ and $\phi=0$. In (a) the vertical axis measures time in units of $\tau =2\lambda t$. In (b) the vertical axis refers to the amplitude of the probability density. As in Figure~\ref{BG2}, the wave packet propagates by oscillating back and forth between the origin and the outer turning point. In this case the probability density is single peaked and spreads as its center approaches the outer turning point. The density profiles at the right correspond to the times  $\tau = 0$ (continuous red curve) , $\tau = \frac{\pi}{2}$ (dotted black curve) and $\tau = \pi$ (dashed blue curve).  
}
\label{P2}
\end{figure}

\vskip1ex
\noindent
$\bullet$ {\bf Time evolution.} The time-evolution of these coherent states is also encoded in the phase of the complex number $z$, namely $\vert z; t \rangle_P = e^{-i\lambda (2\ell+3)t} \vert z(t)\rangle_P$, with $z(t) = \vert z \vert e^{-i(\phi+4\lambda t)}$. Again, the wave packet oscillates back and forth with frequency $2\lambda$ and spreads as its center approximates the outer turning point (see Figure~\ref{P2}).

\vskip1ex
\noindent
$\bullet$ {\bf Uncertainty relation.} As indicated in Section~\ref{algdm}, $L_{\pm}$ correspond to the Schwinger representation of the definite angular momentum ladder operators $A_{\ell}$ and $A_{\ell}^{\dagger}$. In this context, from (\ref{S1a}) we see that $D(\xi)$ in (\ref{operator}) is indeed the two-mode squeeze operator
\be
D(\xi) = e^{\xi a^+ b^+ - \overline\xi b^- a^-}.
\ee
Using (\ref{bosalg1}), (\ref{bosalg2}) and (\ref{S1d}), it is immediate to verify the following transformations
\be
\begin{array}{c}
D(\xi) a^- D^{-1}(\xi)= \cosh (\vert \xi \vert) a^- - \tfrac{\xi}{\vert \xi \vert} \sinh (\vert \xi \vert) b^+ = \widetilde a^-, \\[1ex]
D(\xi) b^+ D^{-1}(\xi)= \cosh (\vert \xi \vert) b^+ - \tfrac{\overline\xi}{\vert \xi \vert} \sinh (\vert \xi \vert) a^- = \widetilde b^+.
\end{array}
\ee
In matrix form we have
\be
U (\xi) \left(
\begin{array}{c}
a^-\\[1ex]
b^+
\end{array}
\right) = \left(
\begin{array}{c}
\widetilde a^-\\[1ex]
\widetilde b^+
\end{array}
\right)
\nonumber
\ee
with $U(\xi) \in SU(1,1)$ given by
\be
U(\xi) = \cosh (\vert \xi \vert) \left(
\begin{array}{cc}
1 & - \tfrac{\xi}{\vert \xi \vert} \tanh (\vert \xi \vert) \\[1ex]
- \tfrac{\overline\xi}{\vert \xi \vert} \tanh (\vert \xi \vert)& 1
\end{array}
\right) = \frac{1}{\sqrt{1-\vert z \vert^2}} \left(
\begin{array}{cc}
1 & - z\\[1ex]
-\overline z& 1 
\end{array}
\right).
\nonumber
\ee
The above expressions are useful in many respects, for instance
\be
D(\xi) L_3 D^{-1}(\xi)=\frac{1}{(1-\vert z \vert^2)} \left[ (1 + \vert z \vert^2)
L_3 -  z L_+ +\overline z L_-
\right]
\ee
leads in natural form to the expectation value
\be
\langle L_3 \rangle_P \equiv {}_P \langle z \vert L_3 \vert z \rangle_P = \frac12 \left(\ell + \frac32 \right) \frac{1 + \vert z \vert^2}{1-\vert z \vert^2}.
\label{ele}
\ee
Then, the average energy of the oscillator is in this case
\be
\overline E^P_{\ell}:= \langle H_{\ell} \rangle_P = 4\langle L_3 \rangle_P = \left(\ell + \tfrac32 \right) \cosh (2 \vert \xi \vert),
\nonumber
\ee
where we have used (\ref{etiq}). The turning points of the time-evolution are the roots of the equation $\overline E^P_{\ell} = V_{\ell}(x)$. Besides (\ref{ele}), the calculation of the following variances is straightforward 
\be
\begin{array}{c}
(\Delta L_1)^2=\frac14 \left( \ell + \frac32\right) \left[1 + \left( \frac{2 \mbox{Re}(z)}{1-\vert z \vert^2} \right)^2 \right], \quad (\Delta L_2)^2=\frac14 \left( \ell + \frac32\right) \left[1 + \left( \frac{2 \mbox{Im}(z)}{1-\vert z \vert^2} \right)^2 \right].
\end{array}
\ee
So that the inequality (\ref{uncer1}) is reduced to an equality only for $z=0$. This last means that only the coherent state $\vert 0 \rangle_P = \vert 0 \rangle_{\ell}$ minimizes the uncertainty between $L_1$ and $L_2$. 

\begin{figure}[htb]

\centering 
\begin{subfigure}[b]{.25\linewidth}
\centering
\includegraphics[width=0.9\textwidth]{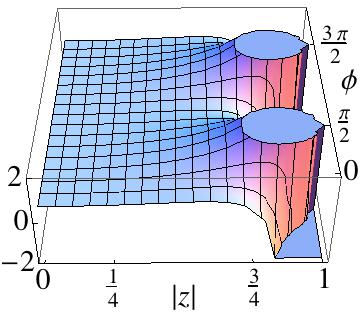} 
\caption{$\frac12 \vert \langle L_3 \rangle_P \vert -(\Delta L_1)^2$}
\end{subfigure}
\hskip3ex
\begin{subfigure}[b]{.25\linewidth}
\centering
\includegraphics[width=0.9\textwidth]{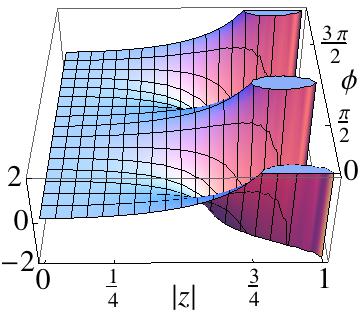}
\caption{$\frac12 \vert \langle L_3 \rangle_P\vert -(\Delta L_2)^2$}
\end{subfigure}

\caption{\footnotesize  Comparison of $\frac12 \vert \langle L_3 \rangle_P \vert$ and $(\Delta L_{1,2})^2$ for the Perelomov coherent states (\ref{mos1}) with $\ell=0$. The nonnegative regions correspond to the fulfilling of the condition $\frac12 \vert \langle L_3 \rangle_P \vert > (\Delta L_{1,2})^2$ for which squeezing occurs. Notice that the squeezing alternates between $L_1$ and $L_2$ depending on the phase $\phi$ of the complex parameter $z$ while $\vert z \vert =0$ produces $\tfrac12 \vert  \langle L_3 \rangle_P \vert = (\Delta L_{1,2})^2$.
}
\label{P3}
\end{figure}

\vskip1ex
\noindent
$\bullet$ {\bf Squeezing and nonclassical behaviour.} For other values of $z$ occurs squeezing if the variance of either $L_1$ or $L_2$ is smaller than the related average uncertainty \cite{Wod85}. Explicitly, either
\be
(\Delta L_1)^2 < \frac12 \vert  \langle L_3 \rangle_P \vert \quad \mbox{or} \quad (\Delta L_2)^2 < \frac12 \vert \langle L_3 \rangle_P \vert
\label{variances}
\ee
is fulfilled. In general, according to (\ref{uncer1}), the variance of one quadrature is reduced at the expense of the other. In Figure~\ref{P3} we can appreciate that the phase $\phi$ alternates such  squeezing between the quadratures (compare with \cite{Ger88}). In particular, $L_1$ is squeezed for $\phi = \frac{\pi}{2}, \frac{3\pi}{2}, \ldots$, and $L_2$ for $\phi =0,\pi,2\pi, \ldots$ Thus, $L_2$ is squeezed around the turning points identified in Figure~\ref{P2} because they are reached by the wave packet center at the times in which $z(t)$ has a global phase equal to entire multiples of $\pi$. In turn, $L_1$ is squeezed at one and three quarters of the entire period. Similar properties are true for the Perelomov coherent states $\vert z \rangle_P$ associated with other $\ell$-hierarchies.

To be more precise, it is well known that the Glauber-Sudarshan $P$-function \cite{Gla07,Sud63} used to calculate a given variance $(\Delta X)^2$ that is squeezed  does not have the character of a probability distribution  because it is either negative or highly singular \cite{Wal79}. Therefore, the related system has not classical description. This is precisely the case of the quadrature variances (\ref{variances}), so that the radial oscillator prepared in the coherent state $\vert z \rangle_P$ is nonclassical for $z \neq 0$ and $\phi = n \tfrac{\pi}{2}$, $n=0,1,2,\ldots$

\subsection{Spectrum generating algebra of definite energy hierarchies}
\label{algde}

\begin{figure}[htb]
\centering 
\includegraphics[width=0.45\textwidth]{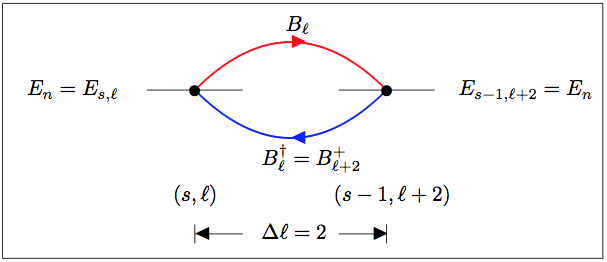}

\caption{\footnotesize Action of the definite energy ladder operators $B_{\ell}$ and $B^{\dagger}_{\ell} =B^+_{\ell+2}$. In the energy space configuration they preserve the energy eigenvalue but increase and decrease the angular momentum eigenvalue $\ell$ in two units respectively. In the $(\ell,s)$-plane, the action of $B_{\ell}$ ($B^{\dagger}_{\ell}$ ) annihilates (creates) a node in the position representation of the definite energy vector $\vert s, \ell\rangle$.
}
\label{Bops}
\end{figure}

Using the upper and lower compositions of left and right arrows in Figure~\ref{enconf} we obtain  the operators
\be
B_{\ell} =a_{\ell+1} b^{\dagger}_{\ell+1} = b^{\dagger}_{\ell+2} a_{\ell}, \qquad
B^+_{\ell}=b_{\ell-1} a^{\dagger}_{\ell-1} = a^{\dagger}_{\ell-2} b_{\ell}, \qquad B_{\ell}^{\dagger} \neq B^+_{\ell}.
\label{ladB}
\ee
In contrast with $A_{\ell}$ and $A^{\dagger}_{\ell}$, the operators (\ref{ladB}) modify both parameters in the ket  $\vert s, \ell \rangle$. That is
\be
B_{\ell} : (\ell, s)  \mapsto (\ell +2, s-1), \qquad B^+_{\ell} : (\ell, s)  \mapsto (\ell -2, s+1).
\nonumber
\ee
This last is the reason for which $B_{\ell}^{\dagger} \neq B^+_{\ell}$. Indeed, the adjoint conjugate of $B_{\ell}$ is $B^+_{\ell+2}$ rather than $B^+_{\ell}$. In this form, for any $s'\geq 0$, 
\be
B^{\dagger}_{\ell}=B^+_{\ell+2} : (\ell+2, s')  \mapsto (\ell, s'+1).
\nonumber
\ee
Thus, $B_{\ell}$ and $B^+_{\ell}$ operate in a specific $E_n$-hierarchy by decreasing and increasing the radial quantum number $s$ in one unit (see Figure~\ref{Bops}). 

The straightforward calculation shows that the following intertwining relationships hold
\be
B_{\ell} H_{\ell} = H_{\ell+2} B_{\ell}, \qquad B^+_{\ell} H_{\ell} = H_{\ell -2} B^+_{\ell}.
\label{intB}
\ee
Therefore, 
\be
[B_{\ell}, H_{\ell}]= \frac{2(2\ell+3)}{r^2} B_{\ell}, \quad [B^+_{\ell}, H_{\ell}]=- \frac{2(2\ell-1)}{r^2} B^+_{\ell},
\nonumber
\ee
where we have used
\be
H_{\ell+2} = H_{\ell} +\frac{2(2\ell+3)}{r^2} , \qquad H_{\ell-2} = H_{\ell} - \frac{2(2\ell-1)}{r^2}.
\nonumber
\ee
It is convenient to rewrite (\ref{intB}) as
\be
B_{\ell} H_{\ell} = H_{\ell+2} B_{\ell}, \qquad B^{\dagger}_{\ell} H_{\ell+2} = H_{\ell} B^{\dagger}_{\ell}.
\nonumber
\ee
Note that $B^{\dagger}_{\ell} B_{\ell}$ and $B_{\ell} B^{\dagger}_{\ell}$ respectively commute with $H_{\ell}$ and $H_{\ell+2}$. Then $B^{\dagger}_{\ell} B_{\ell} = f(H_{\ell})$ and $B_{\ell} B^{\dagger}_{\ell} =g(H_{\ell+2})$, with $f$ and $g$ smooth functions. After some calculations one arrives at the expression
\be
[B^{\dagger}_{\ell}, B_{\ell}] = -8(2\ell +1) \mathbb I \equiv 4(a_{\ell-1} a_{\ell -1}^{\dagger} - b_{\ell} b^{\dagger}_{\ell}) =8D_{\ell}.
\label{su20}
\ee
This last result gives rise to the commutation rules
\be
[D_{\ell}, B^{\dagger}_{\ell}]=4 B^{\dagger}_{\ell}, \qquad [D_{\ell}, B_{\ell}]=-4 B_{\ell}.
\label{su21}
\ee
Following the steps of the previous sections we now introduce the free-index operators\footnote{From (\ref{ladB}) it is clear that one has an additional representation for $J_{\pm}$. Namely, $J_+ =b^- a^+$ and $J_-=a^- b^+$.}: 
\be
J_+=a^+ b^-, \quad J_- = b^+ a^-, \quad J_3 =\tfrac12 (N_s -N_{\ell}),
\label{su22}
\ee
that is
\be
J_+ \leftrightarrow \tfrac14 B^{\dagger}_{\ell}, \qquad J_- \leftrightarrow \tfrac14 B_{\ell}, \qquad J_3  \leftrightarrow \tfrac14 D_{\ell}.
\nonumber
\ee
Then the commutators (\ref{su20}) and (\ref{su21}) correspond to the Lie algebra of $SU(2)$, 
\be
[J_-,J_+]= - 2J_3, \quad [J_3,J_{\pm}]=\pm J_{\pm}.
\label{su24}
\ee
The expressions (\ref{su20})--(\ref{su24}) are the Schwinger representation of the $su(2)$ algebra for which the  definite energy ladder operators $J_{\pm}$ are linked to the two bosons $a^{\pm}$ and $b^{\pm}$, provided that $2J_3 = N_s - N_{\ell}$. In this case the boson occupation $2J_3$ leads to $-(\ell + \tfrac12)$, so that the representation is determined by the orbital quantum number $\ell$.

In addition to the commutation rules (\ref{bosalg1}), (\ref{bosalg2}) and (\ref{S1d}), we have
\be
\begin{array}{c}
[J_{\pm}, a^{\pm}]=[J_{\pm}, b^{\mp}]=0,\\[1ex]
[J_{\pm}, a^{\mp}]=\mp b^{\mp}, \quad [J_{\pm}, b^{\pm}]=\pm a^{\pm}, \quad [J_3, a^{\pm}]= \pm \tfrac12 a^{\pm}, \quad [J_3, b^{\mp}] = \pm \tfrac12 b^{\mp}.
\end{array}
\label{su25}
\ee
The action of $J_{\pm}$ on the vectors belonging to the $E_n$-hierarchy is as follows
\be
J_- \vert n, \ell \rangle_e = \tfrac12\sqrt{(n-\ell)(n+\ell+3)} \vert n, \ell+2\rangle_e, \qquad J_- \vert \ell,\ell \rangle_e =0, 
\label{j-}
\ee
\be
J_+ \vert n, \ell \rangle_e = \tfrac12 \sqrt{(n-\ell+2)(n+\ell+1)} \vert n, \ell-2 \rangle_e.
\label{j+}
\ee
For completeness, in $(\ell,s)$-configuration, the above expressions are written as
\bea
J_- \vert s, \ell \rangle = \sqrt{s(s+\ell+3/2)} \vert s-1, \ell+2\rangle, \qquad J_- \vert 0,\ell \rangle =0,
\nonumber\\[1ex] 
J_+ \vert s, \ell \rangle = \sqrt{(s+1)(s+\ell+1/2)} \vert s+1, \ell-2 \rangle.
\nonumber
\eea
Thus, acting on the points of the $(\ell,s)$-plane, $J_-(J_+)$ produces horizontal displacements by decreasing (increasing) the radial quantum number $s$ in one unit, as desired. At the same time the orbital quantum number $\ell$ is increased (decreased) in two units. In this form, the representation determined by $s$ ($\ell$) implies that $J_-$ and $J_+$ are annihilation (creation) and creation (annihilation) operators respectively. This two-fold profile of $J_{\pm}$ plays a relevant role in the following calculations.

\subsubsection{Representation spaces}
\label{repspace}

As indicated in Section~\ref{secfac}, the dimension of the $E_n$-hierarchy ${\cal H}_{(n)}$ is finite and equal to the accidental degeneracy $d=\mbox{deg}({E}_n)$ of the energy eigenvalue $E_n$. In particular, ${\cal H}_{(0)} = \mbox{span}\{ \vert 0, 0 \rangle_e \}$ and ${\cal H}_{(1)} = \mbox{span}\{ \vert 1,1 \rangle_e \}$ are 1-dimensional because $E_0$ and $E_1$ are not degenerate. In general, to get a finite representation of the operators $J_{\pm}$ and $J_3$ derived in the previous section, it is required the existence of two extremal vectors such that $J_- \vert \varphi_{low} \rangle =0$ and $J_+ \vert \varphi_{hig} \rangle=0$, otherwise the representation would be not finite dimensional. From (\ref{j-}), it is clear that $\vert \varphi_{low} \rangle = \vert \ell, \ell \rangle_e $. However, the states that are annihilated by $J_+ = b^- a^+ = a^+ b^-$ are unphysical. Therefore, in order to get a finite representation of the group that rules the dynamics of the definite energy states, we have to look for an additional realization of the $su(2)$ Lie algebra in the $E_n$-hierarchies. 

Bearing in mind that $\vert n,\ell\rangle_e$ is eigenvector of $J_3$ with eigenvalue $m=-\tfrac14 (2\ell+1)$, and that $n= 2s+\ell$ is fixed while $s=0,1, \ldots, d-1$, we find that the admissible values of $\ell$ are equidistant in two units and even (odd) for $n$ even (odd), going  from 0 (1) to $n$. Let us make the transformation $m \rightarrow \mu = m+\mu_0$, with $\mu_0 =\tfrac14 (n+1)$ for $n$ even and $\mu_0 =\tfrac14 (n+2)$ for $n$ odd. That is
\be
\mu =\left\{
\begin{array}{lll}
\frac{n}{4}- \frac{\ell}{2}, & \ell =0, 2, \ldots, n & \mbox{($n$ even)}\\[1ex]
\frac{n-1}{4} - \frac{(\ell -1)}{2}, & \ell =1, 3, \ldots, n & \mbox{($n$ odd)}
\end{array}
\right. 
\nonumber
\ee
In each case  the lowest value of $\ell$ determines the highest weight $j=\frac{d-1}{2}$ of $\mu$:
\be
j =\left\{
\begin{array}{ll}
\frac{n}{4}, & \mbox{$n$ even}\\[1ex]
\frac{n-1}{4}, & \mbox{$n$ odd}
\end{array}
\right.
\label{jotas}
\ee
Therefore
\be
\mu =\left\{
\begin{array}{lll}
j- \frac{\ell}{2}, & \ell =0, 2, \ldots, 4j & \mbox{($n$ even)}\\[1ex]
j - \frac{(\ell -1)}{2}, & \ell =1, 3, \ldots, 4j+1 & \mbox{($n$ odd)}
\end{array}
\right. 
\label{mu1}
\ee
is eigenvalue of the diagonal operator $S_3=\mbox{diag}(j, j-1, \ldots, -j)$. Thus, for $j$ fixed, the  vectors 
\be
\vert n, \ell \rangle_e = \left\{
\begin{array}{ll}
\vert 4j, 2(j-\mu) \rangle_e \equiv \vert j, \mu \rangle_J, & \mbox{$n$ even}\\[1ex]
\vert 4j+1, 2(j-\mu) + 1\rangle_e \equiv \vert j, \mu \rangle_J, & \mbox{$n$ odd}
\end{array}
\right.
\label{mu2}
\ee
are eigenstates of $S_3$ with eigenvalue $\mu$. In other words, the set (\ref{mu1})-(\ref{mu2}) solves the spectral decomposition of $S_3$. 

The space of states of the radial oscillator is in the present case decomposed into the direct sum $\bigoplus_{n=0}^{\infty} {\cal H}_{(n)}$, where the definite energy hierarchies ${\cal H}_{(n)}$ are `horizontal' subspaces of dimension $d=(2j+1)$.

We now proceed to get the representation of the ladder operators $S_{\pm} \in SU(2)$ in the basis (\ref{mu2}). With this aim let us rewrite 
\be
\ell=\left\{
\begin{array}{ll}
2(k-1), \quad k=1,2,\ldots, 2j+1 & (\mbox{$n$ even})\\[1ex]
2k-1, \quad k=1,2,\ldots, 2j+1 & (\mbox{$n$ odd})
\end{array}
\right.
\nonumber
\ee
so that $\mu \leftrightarrow \mu_p =j-p+1$ with $p=1,2,\ldots, 2j+1$, regardless of wheter $n$ is even or odd. Accordingly, the relationship (\ref{mu2}) is simplified
\be
\vert j, \mu_p \rangle_J = \left\{
\begin{array}{ll}
\vert n, 2(p-1) \rangle_e,  & \mbox{$n$ even}\\[1ex]
\vert n, 2p-1 \rangle_e, & \mbox{$n$ odd}
\end{array}
\right. ; \quad p=1,2,\ldots, 2j+1.
\label{mu3}
\ee
In particular, the extremal states $\vert j, j \rangle_J$ and $\vert j, -j \rangle_J$ are obtained for $p=1$ and $p=2j+1$ respectively; they are linked to definite energy states as follows
\be
\vert j, j \rangle_J = \left\{
\begin{array}{l}
\vert n, 0 \rangle_e\\[1ex] 
\vert n, 1 \rangle_e
\end{array}
\right. ; \quad \vert j, -j \rangle_J = \left\{
\begin{array}{l}
\vert n, n \rangle_e \\[1ex] 
\vert n, n \rangle_e
\end{array}
\right. \qquad
\begin{array}{l}
(\mbox{$n$ even}) \\[1ex] 
 (\mbox{$n$ odd})
\end{array}
\label{mu4}
\ee
Now, according to \cite{Enr13,Enr14}, let us introduce the dyadic expressions
\be
X_{2j+1}^{p,q}:= \vert j, \mu_p \rangle_J \langle j, \mu_q \vert \equiv \left\{
\begin{array}{ll}
\vert n, 2(p-1) \rangle_e \langle n, 2(q-1) \vert, & \mbox{$n$ even}\\[1ex]
\vert n, 2p-1\rangle_e \langle n, 2q-1 \vert, & \mbox{$n$ odd}
\end{array}
\right.
\nonumber
\ee
which are Hubbard operators in their simplest representation and correspond to $(2j+1)$-square matrices that have entry 1 in position $(p,q)$ and zero in all other entries. In this form we have the Hubbard representation of the operators we are looking for
\be
\begin{array}{c}
S_3 = \displaystyle\sum_{k=1}^{d} \mu_k X_d^{k,k}, \\[3ex]
S_+ = \displaystyle\sum_{k=1}^{d -1} \sqrt{k(2j+1-k)} X_d^{k, k+1}, \quad S_- = \sum_{k=1}^{d -1} \sqrt{k(2j+1-k)} X_d^{k+1, k}.
\end{array}
\ee
These last generate the $su(2)$  Lie algebra in the space of definite energy states
\be
[S_3, S_{\pm}] = \pm S_{\pm}, \quad [S_-, S_+]= -2S_3.
\label{opS}
\ee
We have shown that $SU(2)$ is the generating group of the definite energy hierarchies. Here the spectrum is the set of values that can take either the radial quantum number $s$ or the orbital one $\ell$, provided that $n$ is fixed.

The action of $S_3$ and $S_{\pm}$ on the elements of the $E_n$-hierarchy is as follows
\be
\begin{array}{c}
S_3 \vert j, \mu_p \rangle_J = \mu_p \vert j, \mu_p \rangle_J, \\[1.5ex]
S_+ \vert j, \mu_p \rangle_J = \sqrt{(p-1)(2j+2-p)} \vert j, \mu_{p-1} \rangle_J,\\[1.5ex] 
S_- \vert j, \mu_p \rangle_J = \sqrt{p(2j+1-p)} \vert j, \mu_{p+1} \rangle_J.
\end{array}
\ee
Iterating the action of $S_+$ we arrive at the vector
\be
S^k_+ \vert j, \mu_p \rangle_J = \sqrt{ \frac{\Gamma(p) \Gamma(2j+k+2-p)}{\Gamma(p-k) \Gamma(2j+2-p)}} \vert j, \mu_{p-k} \rangle_J.
\ee
In particular, for $p=2j+1$ the above equation leads to
\be
S^k_+ \vert j, -j \rangle_J = \sqrt{\frac{ \Gamma(2j+1) \Gamma(k+1)}{\Gamma(2j+1-k)}} \vert j, -j+k \rangle_J.
\label{pcs1}
\ee
Besides, for $k=2j$ one gets $S_+^{2j} \vert j,-j\rangle_J = \Gamma(2j+1) \vert j, j \rangle$, so that $S_+^{2j+1} \vert j,-j\rangle_J =0$. Therefore, given $z \in \mathbb C$,
\be
e^{zS_+} \vert j, -j \rangle_J =\sum_{k=0}^{2j} \left[\frac{ \Gamma(2j+1)}{\Gamma(2j+1-k) \Gamma(k+1)} \right]^{1/2} z^k \vert j, -j+k \rangle_J.
\label{pcs2}
\ee

\subsubsection{$SU(2)$ Perelomov coherent states}
\label{sec322}

We now use $\vert j, -j \rangle_J$ as the extremal state to be displaced in order to get the related Perelomov coherent states. That is,
\be
\vert n, z \rangle_P =D(\xi) \vert j, -j \rangle_J \quad \mbox{with} \quad D(\xi) = e^{\xi S_+ -\overline \xi S_-} , \quad \xi \in \mathbb C.
\ee
Using the disentangling formula \cite{Per86},
\be
e^{\xi S_+ -\overline \xi S_-} = e^{zS_+} e^{\ln (1+\vert z \vert^2) J_3} e^{-\overline z S_-}, \quad z= \tfrac{\xi}{\vert \xi \vert} \tan (\vert \xi \vert) \in \mathbb C,
\ee
together with (\ref{pcs2}) and the fact that $\vert j, -j \rangle_J$ is annihilated by $S_-$, it is immediate to obtain
\be
\vert n, z \rangle_P = (1+\vert z \vert^2)^{-j} \sum_{k=0}^{2j} \left[\frac{ \Gamma(2j+1)}{\Gamma(2j+1-k) \Gamma(k+1)} \right]^{1/2} z^k \vert j, -j+k \rangle_J.
\label{pcs3}
\ee
Considering (\ref{mu3}) and (\ref{mu4}), in energy space configuration we have
\be
\vert n, z \rangle_P = (1+\vert z \vert^2)^{-j} \sum_{k=0}^{2j} \left[\frac{ \Gamma(2j+1)}{\Gamma(2j+1-k) \Gamma(k+1)} \right]^{1/2} z^k \vert n, n-2k \rangle_e,
\label{pcs4}
\ee
where $n$ is either even or odd. 

\noindent
$\bullet$ {\bf Uncertainty relations and squeezing.} In this case the commutators (\ref{opS}) lead to the inequality
\be
\Delta S_1 \Delta S_2 \geq \tfrac12 \vert \langle S_3 \rangle \vert,
\label{ineqS}
\ee
where the quadratures are defined in usual form
\be
S_1 = \frac12 \left(S_+ + S_-\right), \quad S_2 = \frac1{2i} \left(S_+ - S_-\right).
\nonumber
\ee

\begin{figure}[htb]
\centering 
\begin{subfigure}[b]{.3\linewidth}
\centering
\includegraphics[width=0.9\textwidth]{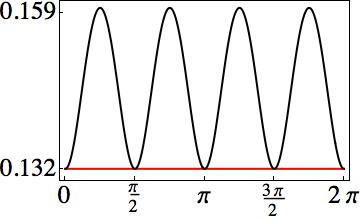} 
\caption{}
\end{subfigure}
\hskip3ex
\begin{subfigure}[b]{.3\linewidth}
\centering
\includegraphics[width=0.9\textwidth]{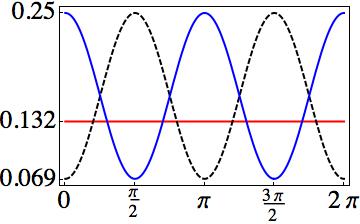}
\caption{}
\end{subfigure}

\caption{\footnotesize The average uncertainty $\tfrac12  \vert \langle S_3 \rangle \vert$, red horizontal line, against (a) the product $\Delta S_1 \Delta S_2$, black continuous curve, and  (b) the variances $(\Delta S_1)^2$ and $(\Delta S_2)^2$, black dashed and blue continuous respectively. In both cases $n=2$ and the horizontal axis measures the phase $\phi$ of the complex parameter $z= \vert z \vert e^{-i\phi}$, with $\vert z \vert =1.8$.
}
\label{Tita}
\end{figure}

\begin{figure}[htb]
\centering 
\includegraphics[width=0.3\textwidth]{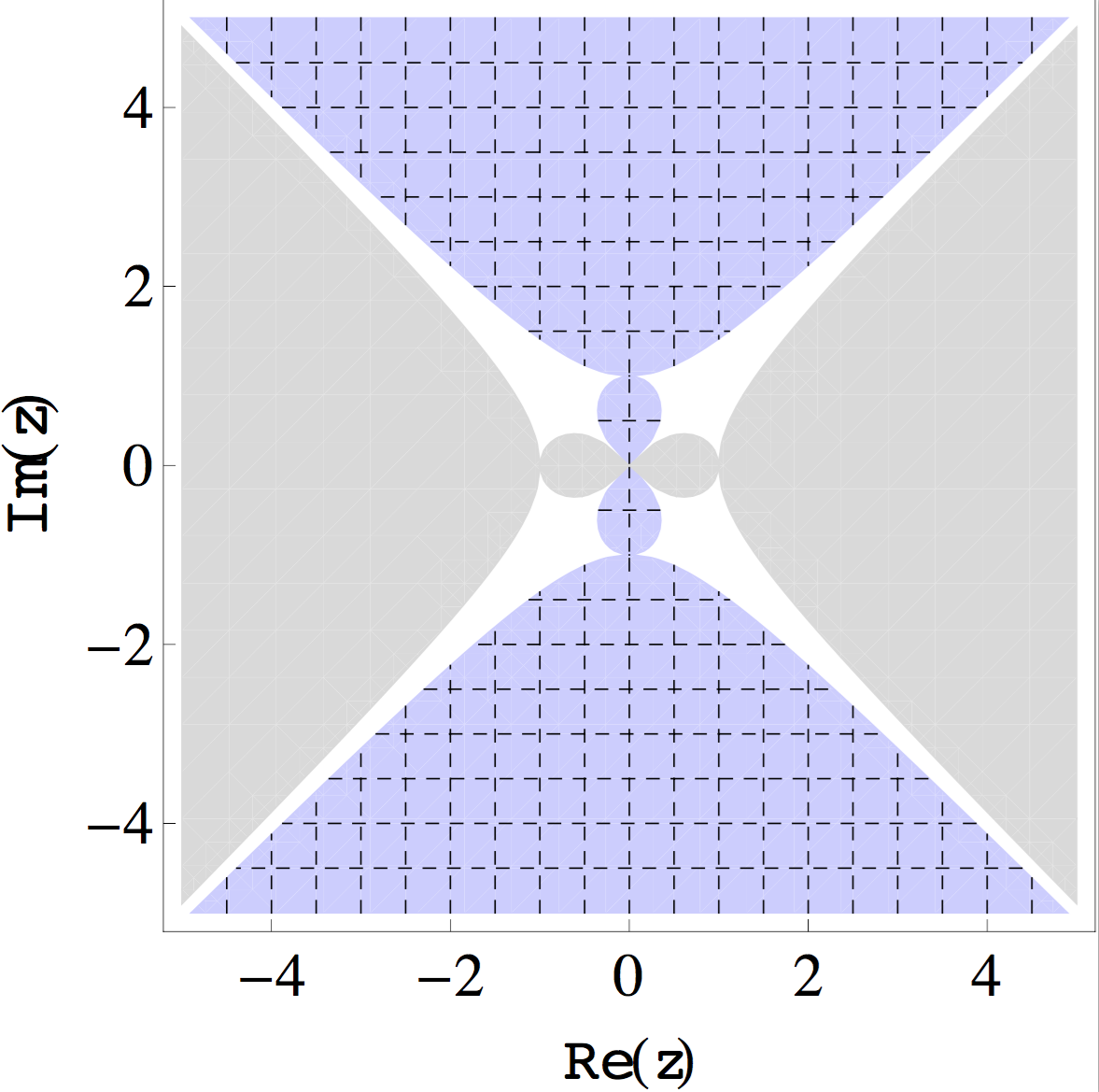} 

\caption{\footnotesize Regions of the complex plane for which squeezing occurs. The grey (mesh, dashed blue) zones correspond to the values of z for which $S_1$ ($S_2$) is squeezed with respect to the corresponding mean value of $S_3$. The white zones indicate the values of $z$ for which both variances are bigger than the average uncertainty. Note that the white zones are centered at $\phi = \tfrac{k\pi}{4}$, $k=1,3,5,7$, while the grey and blue zones are centered at $\phi=0,\pi$ and $\phi=\tfrac{\pi}{2}, \tfrac{3\pi}{2}$ respectively. For $\vert z \vert$ given, there will be classical-to-quantum transitions (and viceversa) as $\phi$ runs from 0 to $2\pi$.
}
\label{Tita2}
\end{figure}

\noindent
After some calculations one gets
\be  
\langle S_3 \rangle_P \equiv\; \!_P\langle n,z \vert S_3 \vert n,z \rangle_P = \frac{1 - \vert z \vert^2}{1 + \vert z \vert^2} \langle S_3 \rangle_0,
\ee
and 
\begin{equation}\label{varS}
  (\Delta S_1)^2 = \frac12 \left[\left(\frac{2 \mathrm{Re}(z)}{1 + \vert z \vert^2}\right)^2 - 1\right] \langle S_3 \rangle_0,
  \quad 
  (\Delta S_2)^2 = \frac12 \left[\left(\frac{2 \mathrm{Im}(z)}{1 + \vert z \vert^2}\right)^2 - 1\right] \langle S_3 \rangle_0,
\end{equation}
with
\be
\langle S_3 \rangle_0 \equiv\; \!_J\langle j,-j \vert S_3 \vert j,-j \rangle_J = -\left\{
\begin{array}{ll}
\frac n4, & n \; \mathrm{even} \\
\frac{n-1}4, & n \; \mathrm{odd} 
\end{array}
\right.
\ee                                                                                 
The behaviour of the above quantities is shown in Figure~\ref{Tita}. We can appreciate that inequality (\ref{ineqS}) becomes an identity for either $\vert z \vert =0$ and any value of $\phi$,  or for $\phi =n\pi/2$, $n=0,1,\ldots$, and any value of $\vert z \vert$ (although this last has been taken equal to 1.8 in the figure). Squeezing of $S_1$ and $S_2$ occurs in the vicinity of $\phi = n\pi$ and $\phi= (2n+1)\frac{\pi}{2}$ respectively (compare with \cite{Wod85}). Thus, we have classical-to-quantum and quantum-to-classical transitions due to the phase of the complex parameter $\vert z \vert e^{-i\phi}$ that do not depend on the value of $\langle S_3 \rangle_0$, see Figure~\ref{Tita2}. As in the squeezing occurring for the definite angular momentum states, one of the quadratures of the definite energy states is reduced at the expense of the other around the above indicated points. Notice however that this is not true for all the values of the phase $\phi$. For instance, taking $\phi\approx \tfrac{\pi}{4}, \tfrac{3\pi}{4}, \ldots$, both of the variances are bigger than the average uncertainty, so that neither squeezing nor minimized uncertainty is present.

\subsection{Transition probabilities}
\label{sectran}

\begin{figure}[htb]
\centering 
\includegraphics[width=0.5\textwidth]{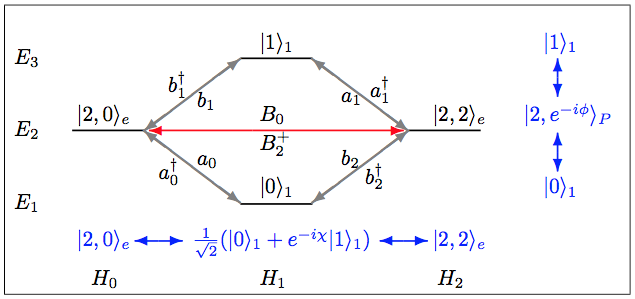}

\caption{\footnotesize Intertwining between the definite energy states $\vert 2,0 \rangle_e$ and $\vert 2,2 \rangle_e$ that share the eigenvalue $E_2 =7$. The transitions $\vert 2,0 \rangle_e \leftrightarrow \vert 2,2 \rangle_e$ are regulated by an intermediary state $\tfrac{1}{\sqrt 2} (\vert 0  \rangle_1 + e^{-i\chi} \vert 1 \rangle_1)$ of undefined energy but definite orbital angular momentum ($\ell=1$). The same diagram corresponds to the transitions $\vert 0 \rangle_1 \leftrightarrow \vert 1 \rangle_1$, these last regulated by an intermediary state $\vert 2, e^{-i\phi} \rangle_P$ of undefined orbital angular momentum but definite energy.
}
\label{TProb1}
\end{figure}

In Section~\ref{algde} we have shown that the definite energy ladder operators $B_{\ell}$ and $B_{\ell}^{\dagger}$ act on the states $\vert n, \ell \rangle_e$ by affecting the orbital quantum number $\ell$ in two units. That is, the transitions between the states of a given $E_n$-hierarchy are ruled by the condition $\Delta \ell = \pm 2$ (see for instance Figure~\ref{TProb1}). The $SU(2)$ Perelomov coherent states (\ref{pcs4}) are indeed a linear combination of vectors such that  the orbital label decreases in two units as the sum index increases one unit. Then, one can show that 
\be
{\cal P}_{n,r} (\vert z \vert) = \left[\frac{ \Gamma(2j+1)}{\Gamma(2j+1-r) \Gamma(r+1)} \right] \frac{\vert z \vert^{2r}}{(1 + \vert z \vert^2)^{2j}} , \quad r=0,1,\ldots,2j,
\nonumber
\ee
is the probability of finding the coherent state with the orbital angular momentum defined by $\ell =n-2r$. However, the Schwinger representation of $B_{\ell}$ and $B_{\ell}^{\dagger}$ includes two bosons, so that the transitions between definite energy states $\vert n, \ell \rangle_e$ must be ruled by the action of the ladder operators $a^{\pm}$ and $b^{\pm}$. In this respect, we should notice that there is a series of intermediary states that connect two arbitrary vectors in a given $E_n$-hierarchy. The intermediary states are by necessity of undefined energy but definite angular orbital momentum. For instance, the ${\cal H}_{(2)}$ hierarchy depicted in Figure~\ref{TProb1} includes only two allowed transitions: $\vert 2,0 \rangle_e \leftrightarrow \vert 2,2 \rangle_e$. Using the boson operators $a^{\pm}$ and $b^{\pm}$ we see that an arbitrary superposition of the states $\vert 0 \rangle_1$ and $\vert 1 \rangle_1$ is involved. As no preference is a priori considered, both intermediary vectors must occur with equal probability, so that we have the transitions
\be
\vert 2,0 \rangle_e \leftrightarrow \tfrac{1}{\sqrt 2} ( \vert 0  \rangle_1 + e^{-i\chi} \vert 1 \rangle_1)
\leftrightarrow \vert 2, 2 \rangle_e, \quad \chi \in [0, 2\pi).
\label{tres}
\ee

\begin{figure}[htb]
\centering 
\includegraphics[width=0.5\textwidth]{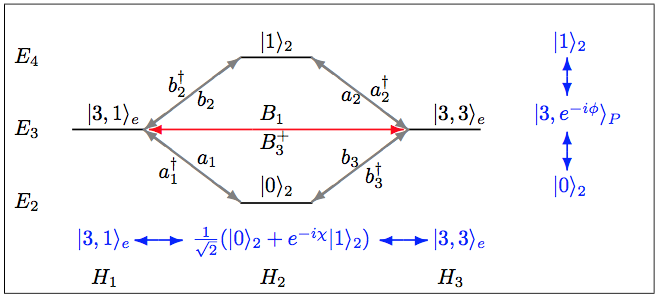}

\caption{\footnotesize Intertwining between the definite energy states $\vert 3,1 \rangle_e$ and $\vert 3,3 \rangle_e$ that share the eigenvalue $E_3 =9$. The transitions $\vert 3,1 \rangle_e \leftrightarrow \vert 3,3 \rangle_e$ are regulated by an intermediary state $\tfrac{1}{\sqrt 2}(\vert 0  \rangle_2 + e^{-i\chi} \vert 1 \rangle_2)$ of undefined energy but definite orbital angular momentum ($\ell=2$). The same diagram corresponds to the transitions $\vert 0 \rangle_2 \leftrightarrow \vert 1 \rangle_2$, these last regulated by an intermediary state $\vert 3, e^{-i\phi} \rangle_P$ of undefined orbital angular momentum but definite energy. Compare with Figure~\ref{TProb1}.
}
\label{TProb2}
\end{figure}

Notice that each one of the three vectors in (\ref{tres}) represents a state of well-defined angular momentum. Moreover, they can be used as the orthonormal basis of a 3-dimensional representation space associated with the $E_2$-hierarchy:
\be
\vert 1,-1\rangle_{D1} = \vert 2, 2\rangle_e, \quad \vert 1,0\rangle_{D1} =\tfrac1{\sqrt 2} (\vert 1, 1\rangle_e+\vert 3, 1\rangle_e),\quad \vert 1,1\rangle_{D1} = \vert 2, 0\rangle_e,
\ee
where $\vert 0 \rangle_1 = \vert 1,1 \rangle_e$, $\vert 1 \rangle_1 = \vert 3,1\rangle_e$, and the phase $\chi$ has been fixed as $\chi=0$ for simplicity. A similar description holds by analyzing the transition between the definite angular momentum states $\vert 0 \rangle_1$ and $\vert 1 \rangle_1$ showed in Figure~\ref{TProb1}. In this case the Perelomov vector $\vert 2, e^{-i\phi} \rangle_P$ plays the role of intermediary state with identical probability of occurrence for the related definite energy states
\be
\vert 1 \rangle_1 \leftrightarrow \tfrac{1}{\sqrt 2} (\vert 2,2 \rangle_e + e^{-i\phi} \vert 2,0 \rangle_e) \leftrightarrow \vert 0 \rangle_1.
\ee
Then, we have a 3-dimensional representation space associated with the $\ell=1$ hierarchy:
\be
 \vert 1,-1\rangle_{D2}=\vert 3, 1\rangle_e, \quad \vert 1,0\rangle_{D2} =\frac1{\sqrt 2} (\vert 2, 0\rangle_e+\vert 2, 2\rangle_e),\quad \vert 1,1\rangle_{D2} = \vert 1, 1\rangle_e,
\ee
where we have taken $\phi=0$. Other two representation spaces of dimension 3 can be achieved by considering the diagram shown in Figure~\ref{TProb2}. For the $E_3$ and $\ell =2$ hierarchies we respectively have
\bea
\vert 1,-1\rangle_{D3} = \vert 3, 3\rangle_e, \quad \vert 1,0\rangle_{D3} =\tfrac1{\sqrt 2} (\vert 4, 2\rangle_e+\vert 2, 2\rangle_e),\quad \vert 1,1\rangle_{D3} = \vert 3, 1\rangle_e, \\[1ex]
 \vert 1,-1\rangle_{D4}=\vert 4, 2\rangle_e, \quad \vert 1,0\rangle_{D4} =\tfrac1{\sqrt 2} (\vert 3, 1\rangle_e+\vert 3, 3\rangle_e),\quad \vert 1,1\rangle_{D4} = \vert 2,2\rangle_e,
\eea
where $\vert 0 \rangle_2 = \vert 2,2 \rangle_e$ and $\vert 1 \rangle_2 = \vert 4,2\rangle_e$, with $\chi=\phi=0$.

\subsubsection{Dicke-like states}
\label{sec331}

We have seen that the transitions between the physical states of the 2-dimensional hierarchies $E_2$ and $E_3$ require an intermediary state of undefined energy. Thus, the transitions $H_{\ell} \rightarrow H_{\ell \pm 1}$, $\ell=1,2$, allowed by the intertwining relationships defined in the previous sections, demand a representation space bigger than ${\cal H}_{(2)}$. Indeed, the four cases discussed above are different representation spaces of dimension 3 for the $SU(2)$ group. All of them correspond to the angular momentum that is defined by the highest weight $j_{D_k}=1$, this last justifies the notation $\vert j_{D_k}, \mu_{D_k} \rangle_{D_k}$, with $\mu_{D_k} = 1, 0, -1$, and $k=1,2,3,4$.

\begin{figure}[th]
\centering 

\includegraphics[width=0.65\textwidth]{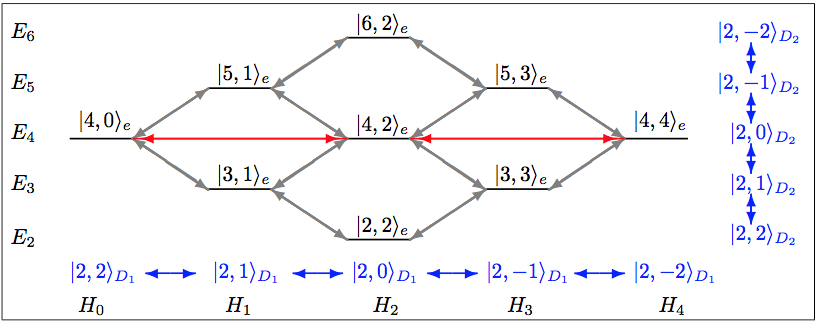}

\caption{\footnotesize Intertwining between the definite energy states $\vert 4,0 \rangle_e$, $\vert 4,2 \rangle_e$ and $\vert 4,4 \rangle_e$ that share the eigenvalue $E_4 =11$. The transitions between these vectors are regulated by the intermediary states $\vert 2, k \rangle_{D_1}$, $k=1,0,-1$. The same diagram corresponds to the transitions between the states indicated at the right column of the figure.
}
\label{TProb3}
\end{figure}

In general, given any $E_n$-hierarchy, one has
\be
 j_{Dk}=\left\{ \begin{array}{ll} n/2, & n~{\rm even}\\[1ex] (n-1)/2, & n~{\rm odd}\end{array}\right.
 \nonumber
\ee
Then, we can construct a representation space ${\cal H}^D_{(2j_D +1)}$ of vectors $\vert j_{D_k}, \mu_{D_k} \rangle_{D_k}$ by following the steps indicated in Section~\ref{repspace}. As an example consider the diagram shown in Figure~\ref{TProb3}, where we have omitted the representation of the corresponding ladder operators. The $E_4$-hierarchy is spanned by the orthonormal basis $\vert 4, \ell \rangle_e$, with $\ell =0,2,4$. We have three possible transitions:
\be
\vert 4, 0 \rangle_e \leftrightarrow \vert 4, 2 \rangle_e, \quad \vert 4, 2 \rangle_e \leftrightarrow \vert 4, 4 \rangle_e, \quad \vert 4, 0 \rangle_e \leftrightarrow \vert 4, 4 \rangle_e.
\label{transR}
\ee
The first two require only one intermediary state, that is,
\be
\vert 4, 0 \rangle_e \leftrightarrow \vert 2,1 \rangle_{D_1} \leftrightarrow \vert 4, 2 \rangle_e, 
\qquad \vert 4, 2 \rangle_e \leftrightarrow \vert 2,-1 \rangle_{D_1} \leftrightarrow \vert 4, 4 \rangle_e,
\ee
where 
\be
\vert 2,1 \rangle_{D_1} = \tfrac{1}{\sqrt 2} (\vert 3, 1\rangle_e + \vert 5,1 \rangle_e), \qquad
\vert 2,-1 \rangle_{D_1} = \tfrac{1}{\sqrt 2} (\vert 3, 3\rangle_e + \vert 5,3 \rangle_e).
\ee
The third transition in (\ref{transR}) requires these last two intermediary states and a third one:
\be
\vert 2,0\rangle_{D1} =\tfrac1{\sqrt 3} (\vert 2,2\rangle_e+\vert4,2\rangle_e+\vert6,2\rangle_e).
\ee
The identification $\vert 4,0 \rangle_e = \vert 2,2 \rangle_{D_1}$, $\vert 4,4 \rangle_e = \vert 2,-2 \rangle_{D_1}$ completes the basis $\vert j_{D_1}, \mu_{D_1} \rangle_{D_1}$ for $j_{D_1} =2$. A second representation is obtained by considering the transitions $\vert 6,2 \rangle_e \leftrightarrow \vert 2,2 \rangle_e$ indicated at the right column of Figure~\ref{TProb3}. In addition, other two representations of dimension 5 are associated with the $E_5$-hierarchy.

As in the previous cases, the representation of the generators of the $su(2)$ Lie algebra is constructed using the Hubbard operators 
\be
 X_{n_k}^{p,q}:= \vert j_{D_k},j_{D_k}-p+1\rangle_{D_k} \langle j_{D_k},j_{D_k}-q+1\vert, \quad n_k=2j_{D_k}+1.
 \nonumber
\ee
Then
\be
J^D_3=\sum_{p=1}^{n_k} (j_{D_k}-p+1) X_{n_k}^{p,p}, \quad J^D_+=\displaystyle\sum_{p=1}^{n_k-1} \sqrt{p(2j_{D_k}+1-p)}X_{n_k}^{p,p+1}, \quad J^D_- = (J^D_+)^{\dagger}.
\ee

On the other hand, we say that the vectors $\vert j_{D_k}, \mu_{D_k} \rangle_{D_k}$ are {\em Dicke-like} states in resemblance with the permutation invariant states that describe a system of $n$ qubits with $k$ components in the excited state \cite{Dic54}. In the canonical basis, the Dicke states are of the form
\be
\vert D(n,k)\rangle = \binom n k^{-1/2} \sum_{\pi\in S_n} \vert\pi( \underbrace{0\cdots0}_k\underbrace{1\cdots 1}_{n-k})\rangle, \quad\binom n k=\frac{n!}{k!(n-k)!}.
\nonumber
\ee
The sum is accomplished over all the permutations of qubits and $S_n$ denotes the symmetric group of order $n$. These states constitute an irreducible representation of $SU(2)$ defined by the so called {\it collective operators}:
\be
 \hat J_3 = \sum_{i=1}^n \sigma_3^{(i)}, \quad \hat J_\pm = \sum_{i=1}^n \sigma_\pm^{(i)},
 \nonumber
\ee
with $\sigma_{3, \pm}^{(i)}$ acting on the $i$th qubit. Although our model is mono-partite and the Dicke states are defined for multi-partite systems, the following analogy can be established. The vectors $\vert j_{D_k}, \mu_{D_k} \rangle_{D_k}$  are either states of definite angular momentum (horizontal blue expressions in Figures~\ref{TProb1}, \ref{TProb2} and \ref{TProb3}) or states of definite energy (vertical blue expressions in the above quoted figures) likewise to the fact that
the Dicke vectors are states of definite number of excitations. Therefore, we can consider a multi-qubit system as follows. 

Figure~\ref{TProb1} is also the representation of two qubits, the first one is of definite orbital quantum number $\ell=1$ and admits only one of two possible values of the energy, $E_3$ and $E_1$ (the zero of the energy is at $E_2$). The second qubit is of definite energy $E_2=7$ and admits only one of two possible values of the orbital quantum number, $\ell=0$ and $\ell =2$ (the zero is at $\ell=1$). Thus, the diagram shown in Figure~\ref{TProb1} corresponds to a bi-partite qubit system of radial oscillators ${\cal S} = {\cal S}_A + {\cal S}_B$, where ${\cal S}_A$ is characterized by the energy and  ${\cal S}_B$ by the orbital quantum number. In the canonical basis we can write $\vert 3,1 \rangle_e = \vert 0 \rangle_A$, $\vert 1,1 \rangle_e = \vert 1 \rangle_A$, $\vert 2,0 \rangle_e = \vert 1 \rangle_B$, and $\vert 2,2 \rangle_e = \vert 0 \rangle_B$. Then, the most general state of the full system ${\cal S}$ acquires the form
\be
\begin{array}{rl}
\vert \Psi \rangle  & =   a \vert 0 \rangle_A \otimes \vert 0 \rangle_B + b \vert 0 \rangle_A \otimes \vert 1 \rangle_B + c \vert 1 \rangle_A \otimes \vert 0 \rangle_B + d \vert 1 \rangle_A \otimes \vert 1 \rangle_B\\[1ex]
& =  a \vert 3,1 \rangle_e \otimes \vert 2,2 \rangle_e + b  \vert 3,1 \rangle_e \otimes \vert 2,0 \rangle_e + c \vert 1,1 \rangle_e \otimes \vert 2,2 \rangle_e + d  \vert 1,1 \rangle_e \otimes \vert 2,0 \rangle_e,
\end{array}
\nonumber
\ee
where $a,b,c$ and $d$ are such that $\vert \Psi \rangle$ is normalized. The concurrence $C(\vert \Psi \rangle) = 2 \vert ad -bc \vert$, $0 \leq C \leq 1$, is a measure of the entanglement \cite{Hil97}  between the two radial oscillators ${\cal S}_{A,B}$. For instance, taking $c=d=0$ we have $C=0$ (separable) while for $a=d=0$, $b=c =\tfrac{1}{\sqrt 2}$,  we have $C=1$ (maximally entangled). Applying the appropriate interactions the transitions showed in Figure~\ref{TProb1} should correspond to the coherences of the related density operator (see, e.g. \cite{Enr14b,Qui15,Qui16}). Other possibility for the intermediary states is to take them in correspondence with the coherent and squeezed states discussed in the previous section. Further details will be reported elsewhere.

A similar description is true for the elements in Figure~\ref{TProb2}. For vector spaces ${\cal H}^D_{(2j_D +1)}$ of higher dimension (as the one associated with Figure~\ref{TProb3}), the number of qubits increases as the value of $j_{D_k}$.

\section{Spectrum generating algebra of `diagonal' hierarchies }
\label{secdiagonal}

\begin{figure}[htb]
\centering 
\includegraphics[width=0.5\textwidth]{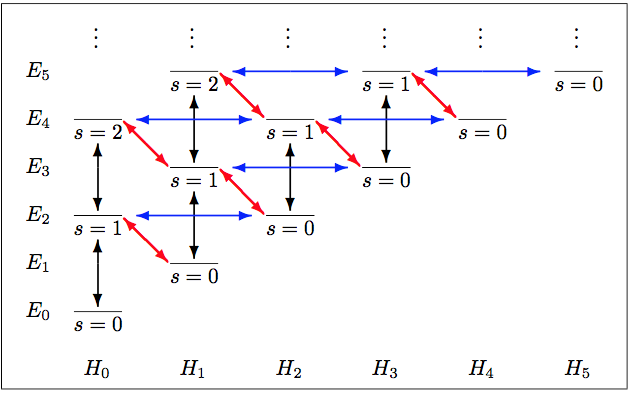}

\caption{\footnotesize Ladder operators in the energy space configuration. The definite angular momentum ladder operators $A_{\ell}$ and $A^{\dagger}_{\ell}$, black--vertical arrows, operate in a specific $\ell$-hierarchy of state vectors by annihilating and creating the radial quantum number $s$ respectively. In turn, the energy definite ladder operators $B_{\ell}$ and $B^{\dagger}_{\ell}$, blue--horizontal arrows, operate in a specific $E_n$-hierarchy of state vectors by annihilating and creating the radial quantum number $s$ respectively. A third type of ladder operators, namely $C_{\ell}$ and $C^{\dagger}_{\ell}$, operate in the finite--dimensional subspaces spanned by the vectors belonging to the `diagonal' energy levels, as indicated by the red--diagonal arrows. 
}
\label{Cops}
\end{figure}

In the previous sections we have shown that $su(1,1)$ and $su(2)$ are the spectrum generating algebras of the definite angular momentum hierarchies ${\cal H}_{\ell}$ and the definite energy hierarchies ${\cal H}_{(n)}$ respectively. Such algebras are defined by the ladder operators $A_{\ell}$, $A^{\dagger}_{\ell}$, $B_{\ell}$ and $B^{\dagger}_{\ell}$, that modify the radial quantum number $s$ in either vertical or horizontal form, see Figure~\ref{Cops}. There is, however, an additional set of finite-dimensional spaces spanned by the vectors that are intertwined in diagonal form. These last are connected by diagonal red arrows in Figure~\ref{Cops}. The boson operators $a^{\pm}$ defined in (\ref{bosalg1}) seem to be the natural option to produce such an intertwining. Indeed, we introduce the operators $C_3$ and $C_{\pm}$ as
\be
C_3=  \tfrac12 (2 N_s -  N_{\ell} +\tfrac12), \quad C_-= \sqrt{N_{\ell} -N_s -\tfrac12} \, a^-, \quad C_+ = a^+ \sqrt{N_{\ell} -N_s -\tfrac12}.
\ee
They span the Lie algebra of $SU(2)$ because satisfy the commutation rules
\be
\left[C_3, C_{\pm} \right] = \pm C_{\pm}, \quad \left[C_-, C_+\right] = -2 C_3.
\ee

\noindent
The action of this new set of operators on the vectors $\vert s, \ell \rangle$ is given by
\be
\begin{array}{c}
C_3 \vert s, \ell \rangle = \tfrac12 (s-\ell) \vert s, \ell \rangle,\\[2ex]
 C_- \vert s, \ell \rangle = \sqrt{s(\ell + 1)} \; \vert s - 1, \ell + 1 \rangle, \quad C_+ \vert s, \ell \rangle = \sqrt{(s + 1)\ell} \; \vert s + 1, \ell - 1 \rangle,
\end{array}
\ee
with $C_- \vert 0, \ell \rangle =0$ and $C_+ \vert s, 0 \rangle =0$, as desired. In other words, following the steps indicated in Section~\ref{repspace}, $\vert \varphi_{low} \rangle = \vert 0, \ell \rangle$ and  $\vert \varphi_{hig} \rangle = \vert s,0 \rangle$ are the extremal vectors that we require to get a finite representation of $SU(2)$. The highest weight of the eigenvalue $\mu_C = \tfrac12 (s-\ell)$ is reached for the lowest orbital angular momentum,  that is $j_C= \tfrac12 s$. This last means that the representation will be determined by the even values of the principal quantum number because, for $\ell =0$,  we have $n=2s$ with $s$ a nonnegative integer (see the energy levels connected by red arrows in Figure~\ref{Cops}). In this form, 
\be
C_-^k \vert \varphi_{high} \rangle =C_-^k \vert s,0 \rangle = \sqrt{ \frac{\Gamma(s+1) \Gamma(k+1)}{\Gamma(s-k+1)} } \, \vert s-k, k\rangle.
\ee
In particular, for $k=s$ we have $C_-^s \vert s, 0 \rangle = \Gamma(s+1) \vert 0, s \rangle$, so that $C_-^{s+1} \vert s, 0 \rangle =0$. In other words, the dimension of the representation is $d_C = 2j_C+1=s+1$. Thus, for $j_C$ fixed, the vectors
\be
\vert s, \ell \rangle = \vert 2j_C, 2(j_C - \mu_C) \rangle \equiv \vert j,\mu \rangle_C, \quad \mu_C= j_C -\tfrac12 \ell, \quad \ell = 0, 1, \ldots, s,
\ee
are eigenstates of $C_3$ with eigenvalue $\mu_C$. In such a representation we have
\be
\begin{array}{l}
C_3 \vert j, \mu \rangle_C =  \mu_C \vert j, \mu \rangle_C, \\[1ex]
C_- \vert j, \mu \rangle_C =  \sqrt{(j _C + \mu_C )(j _C- \mu_C + 1)} \vert j, \mu - 1 \rangle_C,\\[1ex]
C_+ \vert j, \mu \rangle_C =  \sqrt{(j_C - \mu_C)(j_C + \mu_C + 1)} \vert j, \mu+ 1 \rangle_C.
\end{array}
\label{ces}
\ee
The full space of states of the radial oscillator decomposes into the direct sum $\bigoplus_{d_C=1}^{\infty} {\cal H} ^{d_C}$, where the `diagonal' subspaces are defined as ${\cal H}^{d_C} = \{ \vert s, \ell \rangle, s+\ell =2j_C \}$.

The construction of the $SU(2)$ coherent states can be accomplished as in the previous cases. These have properties similar to the ones discussed for the states of Section~\ref{sec322}.

To conclude this section let us emphasize that the operator multiplying $a^{\pm}$ in the definition of $C_{\pm}$ acts on the elements of ${\cal H}^{d_C}$ as follows
\be
(N_{\ell} -N_s -\tfrac12 ) \vert s, \ell \rangle \equiv \hat \ell_{op}  \vert s, \ell \rangle
= \ell \vert s, \ell \rangle.
\ee
Hence, for $\ell \neq 0$, we can write
\be
C_3 = \frac{1}{2\hat \kappa_{op}} \left( \hat \kappa_{op} N_s - \mathbb I \right), \quad C_- = \frac{1}{\sqrt{-\hat \kappa_{op}}} \,  a^-, \quad C_+ = a^+ \frac{1}{\sqrt{ - \hat \kappa_{op}}}, \quad \hat \kappa_{op} = - \frac{1}{\hat \ell_{op}}.
\ee
Quite recently, in connection with the generalized oscillator algebra reported in \cite{Dao86}, a similar set of operators (with a negative number $\kappa$ instead of the operator $\hat \kappa_{op}$) has been introduced in \cite{Ata10}. There, the authors show that the parameter $\kappa$ defines the dimension of the representation $d=1-1/\kappa$ and gives rise to a phase factor in the relations (\ref{ces}). This last is then shown to be essential in generating mutually unbiased bases. In our case, the algebra spanned by $C_3$ and $C_{\pm}$ is defined in a vector space integrated by states of the radial oscillator that differ in angular momentum and energy as well, so that the operator $\hat \kappa_{op}$ cannot be substituted by a number that is common to all the vectors $\vert s, \ell \rangle$. Indeed, the basis elements of ${\cal H}^{d_C}$ are eigenvectors of $\hat \kappa_{op}$ with non-degenerate  eigenvalue $0> \kappa = -1/\ell$. Notice however that $\vert \varphi_{low} \rangle = \vert 0, \ell \rangle$ is eigenvector of $\hat \kappa_{op}$ with eigenvalue $-1/\ell$. This last defines the dimension of the representation because $d_C = 1-1/\kappa = 1+\ell =1+2j_C$. Further insights on the matter will be reported elsewhere. 

\section{Conclusions}
\label{conclu}

We have used the factorization method to show that  each one of the Lie algebras $su(1,1)$ and $su(2)$ determine the energy spectrum of the radial oscillator. Accordingly, the space of states ${\cal H}$ of the radial oscillator decomposes into the following direct sums:

\begin{itemize}
\item[a)]
${\cal H} = \bigoplus_{\kappa=3/4}^{\infty} {\cal H}^{\kappa}$, with ${\cal H}^{\kappa}$ the infinite-dimensional hierarchies ${\cal H}_{\ell}$ defined in Section~\ref{secfac} and $\kappa= \tfrac14 (2\ell +3) =\tfrac14 \epsilon_{\ell}$ the Bargmann index characterizing each `vertical' subspace ${\cal H}^{\kappa}$ as a representation space of the non-compact Lie group $SU(1,1)$.

\item[b)] 
${\cal H} = \bigoplus_{n=0}^{\infty} {\cal H}_{(n)}$, where the `horizontal' subspaces ${\cal H}_{(n)}$ are the finite-dimensional hierarchies defined in Section~\ref{secfac}, each one a representation space of the compact Lie group $SU(2)$. The standard angular momentum basis $\vert j, \mu \rangle_J \leftrightarrow \vert n, \ell \rangle_e$ is introduced in (\ref{jotas})--(\ref{mu2}), with $j =\tfrac{d-1}{2}$ the highest weight of $\mu$ and $d$ the dimension of ${\cal H}_{(n)}$. 

\item[c)]
${\cal H} = \bigoplus_{d_C=1}^{\infty} {\cal H}^{d_C}$, where each `diagonal' subspace ${\cal H}^{d_C} = \{ \vert s, \ell \rangle, s+\ell=2 j_C \}$ is a representation of $SU(2)$. Here $d_C = 2j_C+1$ is the dimension of the representation and $j_C$ is the highest weight of $\mu_C$.

\item[d)]
Besides the representations (b--c), we have constructed diverse finite-dimensional representations ${\cal H}^D_{2j_d +1}$ of $SU(2)$ that are spanned by the Dicke-like states $\vert j_{D_k}, \mu_{D_k} \rangle_{D_k}$ introduced in Section~\ref{sec331}.

\end{itemize}

\noindent
The emphasis of this work is on $SU(2)$ as the generating group of the radial oscillator because the finite-dimensional representation spaces (b--d) have been unnoticed in the literature on the matter. Another relevant point in our approach is the two-boson (Schwinger) profile of the generators of $su(1,1)$ and $su(2)$ that arises as a natural consequence of applying the factorization method on the radial oscillator Hamiltonian. On the one hand, this Schwinger structure justifies the squeezing observed in the quadrature variances of the coherent states for both generating groups, $SU(1,1)$ and $SU(2)$. On the other hand, the Schwinger profile is  a necessity originated by the allowed transitions of the system. These last properties of the representation theory associated with the generating algebras for the radial oscillator have been not previously reported in the literature.

We have also shown that the majority of the coherent states constructed here are squeezed for the quadratures under the criterion of Wodkiewicz and Eberly \cite{Wod85}. In particular, for the $SU(2)$ Perelomov coherent states constructed in Section~\ref{sec322}, the squeezing depends on the complex parameter $z= \vert z\vert e^{-i\phi}$ that labels the linear combination of definite energy states. For any value of $\vert z \vert$ and $\phi$ in the vicinity of the points $\phi_n = n\tfrac{\pi}{2}$, $n=0,1,2,\ldots$, the squeezing alternates between the two quadratures of the system. The points $\phi_n$ define minimum uncertainty states and the neighbourhoods of $\phi = \tfrac{\pi}{4}, \tfrac{3\pi}{4}, \ldots$, are such that the variances of both quadratures are bigger than the average uncertainty. These classical-to-quantum and quantum-to-classical transitions deserve special attention because they can be controlled by manipulating the phase of $z\in \mathbb C$.

In the literature \cite{Tri94} one can find discussions addressed to motivate the searching of states that satisfy squeezing conditions different from the ones of the Wodkiewicz-Eberly criterion. The main criticism is that for the $SU(1,1)$ coherent states the variances are greater than their value in the ground state. This is certainly true for the states of Sections~\ref{sec311} and \ref{su11per} (see e.g. Figures~\ref{BG3} and \ref{P3}). However, this is not the case for the $SU(2)$ coherent states of Section~\ref{sec322} because the classical-to-quantum and quantum-to-classical transitions do not depend on the ground state of the system (see Figure~\ref{Tita2}). The same holds for the other representations of $SU(2)$ reported in this work. We hope that our results will be useful in the studies on the matter.

Some applications could be found in quantum optics, where the Laguerre-Gaussian modes are useful in finding out the role of the shape in the quality of light beams \cite{Ses02}, and in the comparison of the propagation of light in uniaxial crystals with the propagation in isotropic media \cite{Cin02}. The Gouy phase (one of the photon geometrical phases) has been observed through quantum correlations in Laguerre-Gaussian modes \cite{Kaw08}, the experiment was based on the relative phase of two different modes. The prospects discussed in Section~\ref{sec331} for the Dicke-like states are addressed in this last direction.

\appendix
\section{The 3D isotropic oscillator revisited}
\label{ApA}

\renewcommand{\thesection}{A-\arabic{section}}
\setcounter{section}{0}  

\renewcommand{\theequation}{A-\arabic{equation}}
\setcounter{equation}{0}  
 
The eigenvalue problem associated with the Hamiltonian of the isotropic 3D oscillator 
\be
H= \frac{\mathbf{p}^2}{2m} + \frac12 m\omega^2 \mathbf r^2
\label{ham}
\ee
can be decoupled into the following set of equations
\bea
\label{h}
H \vert n, \ell, m \rangle = {\cal E}_n \vert n, \ell, m \rangle,\\
\label{l2}
\mathbf L^2 \vert n, \ell, m \rangle = \hbar^2 \ell (\ell +1) \vert n, \ell, m \rangle,\\
L_z \vert n, \ell, m \rangle = \hbar m \vert n, \ell, m \rangle,
\label{lz}
\eea
where $\mathbf L$ and $L_z$ are the orbital angular momentum and its projection on the $z$-axis respectively. In position representation, using spherical coordinates $\vert \vec r\rangle= \vert r, \theta, \phi\rangle$, the Hamiltonian (\ref{ham}) acquires the form
\be
H=  \frac{p_r^2}{2m} + \frac{\mathbf{L}^2}{2m r^2} + \frac12 m\omega^2 r^2,
\label{hrad}
\ee
where 
\be
\frac{p_r^2}{2m}= \frac{1}{2m r^2} [(\mathbf r \cdot \mathbf p)^2 -i\hbar (\mathbf r \cdot \mathbf p)] = -\frac{1}{2m} \frac{\hbar^2}{r^2} \frac{\partial}{\partial r} \left( r^2 \frac{\partial}{\partial r} \right)
\nonumber
\ee
corresponds to the {\em radial kinetic} term and 
\be
p_r = \frac{1}{r} (\mathbf r \cdot \mathbf p -i\hbar \mathbb I)
\nonumber
\ee
is the canonical conjugate of the radial position-operator: $[r, p_r ] = i\hbar \mathbb I$. 

It is well known that the solutions of (\ref{l2}) and (\ref{lz}) lead to the (normalized) spherical harmonics
\be
Y_{\ell}^m (\theta, \phi)= (-1)^m \sqrt{\frac{(2\ell +1) \Gamma (\ell -m+1)}{4\pi \Gamma (\ell +m +1)}} P_{\ell}^m(\cos \theta) e^{im\phi}, \quad \ell \geq m\geq 0.
\nonumber
\ee
For $m<0$, it is usual to take $Y_{\ell}^m(\theta,\phi) =(-1)^m Y_{\ell}^{*-m}(\theta,\phi)$, with $z^*$ the complex conjugate of $z \in \mathbb C$.  The set of these functions is orthonormal 
\be
\int_0^{2\pi} \int_0^{\pi}  Y_{\ell}^{*m} (\theta,\phi) Y_{\ell'}^{m'}(\theta,\phi) \sin \theta d\theta d\phi = \delta_{\ell \ell'} \delta_{mm'},
\label{shar}
\ee
where $-\ell \leq m \leq \ell$. Now, from (\ref{hrad}) and (\ref{l2}) we see that the solutions of the eigenvalue equation (\ref{h}) depend only on the variable $r$ and that they should be labelled by $n$ and $\ell$. We have
\be
\frac{\hbar^2}{2m} \left[ -\frac{1}{r^2} \frac{d}{d r} \left( r^2 \frac{d}{d r} \right) +  \frac{\ell (\ell +1)}{r^2} \right]  R_{n\ell}(r) + \frac12 m\omega^2 r^2 R_{n\ell}(r) = {\cal E}_n R_{n\ell}(r),
\nonumber
\ee
with $R_{n\ell}(r) := \langle r \vert n \rangle$. As usual, we take $u_{n\ell}(r) =r R_{n\ell}(r)$ to get the radial equation in standard form
\be
\frac{\hbar^2}{2m}\left[- \frac{d^2}{dr^2} + \frac{\ell (\ell +1)}{r^2} \right] u_{n\ell}(r)  + \frac12 m\omega^2 r^2  u_{n\ell}(r) = {\cal E}_n u_{n\ell}(r).
\nonumber
\ee
Using the dimensionless eigenvalue of the energy $\mathtt{E}_n = \frac{1}{\hbar \omega} {\cal E}_n$ and $\lambda = \frac{m\omega}{\hbar}$,  we arrive at the equation that is analyzed in this work
\be
-\frac{d^2 u_{n\ell}}{dr^2} + \left[ \frac{\ell (\ell+1)}{r^2} +\lambda^2 r^2 -2\lambda \mathtt{E}_n 
\right] u_{n\ell}=0.
\label{eigen}
\ee
In this representation, the resolution of unity
\be
\mathbb I= \int_{\mathbb R^3} \vert \vec r \rangle \langle \vec r \vert d^3 r = \int_0^{\infty} \int_0^{2\pi} \int_0^{\pi}\vert r,\theta,\phi \rangle \langle r,\theta,\phi \vert r^2 \sin \theta d\theta d\phi dr
\nonumber
\ee
leads to
\be
\langle n,\ell,m \vert n', \ell',m'\rangle = \delta_{\ell \ell'} \delta_{mm'} \int_0^{\infty}  R^*_{n\ell}(r) R_{n' \ell}(r) r^2 dr,
\label{inner}
\ee
where we have used the orthonormality of the spherical harmonics (\ref{shar}). Therefore, to get a set of orthonormal vectors $\vert n,\ell, m \rangle$, it follows that the $n$-orthogonal condition 
\be
\int_0^{\infty}  R^*_{n\ell}(r) R_{n' \ell}(r) r^2 dr = \int_0^{\infty}  u^*_{n\ell}(r) u_{n' \ell}(r) dr=\delta_{nn'}
\label{north}
\ee
must be satisfied. We would like to emphasise that (\ref{north}) includes the same $\ell$-index for both $u$-functions because (\ref{shar}) is already fulfilled. That is, the product of two $u$-functions with different $\ell$-index would be different from zero even-though they are labelled with different $n$-indices. Notice however that the integral of $u^*_{n\ell}(r) u_{n' \ell'}(r)$ must be finite in general, otherwise the product (\ref{inner}) is not necessarily zero for $\ell \neq \ell'$.

In this work we pay attention to the states $\vert n, \ell\rangle$ for which the following orthogonality condition is satisfied 
\be
\delta_{nn'} \delta_{\ell \ell'} = \langle n,\ell \vert n',\ell'\rangle= \int_0^{\infty}  u^*_{n\ell}(r) u_{n' \ell}(r) dr \int_0^{\pi} \Theta^*_{\ell m}(\theta) \Theta_{\ell' m}(\theta) \sin \theta d\theta.
\label{inner2}
\ee
Remark that the $\Theta$-functions $\Theta_{\ell m}(\theta):= \langle \theta \vert \ell \rangle$ have the same $m$-index because (\ref{shar}) is already fulfilled. That is, the product (\ref{inner2}) makes sense only for a given eigenvalue of $L_z$. As $m=0$ is included for any value of $\ell$, without loss of generality we take this as the definite value of $m$. In this form the product (\ref{inner2}) is indeed twice indexed. Then, we can rewrite the position representation of the states $\vert n \rangle$ and $\vert \ell \rangle$ associated with the eigenvalue equation (\ref{eigen}) as follows (notice that $\Phi_m(\phi) := \langle \phi \vert m \rangle =1$ for $m=0$):
\be
\langle r\vert n \rangle \equiv u_{n\ell}(r), \qquad \langle \ell \vert \theta \rangle \equiv \Theta_{\ell, m=0}(\theta).
\nonumber
\ee
Bearing in mind this last simplified representation, the radial eigenvalue equation (\ref{eigen}) is expressed as
\be
H_{\ell} \vert n,\ell \rangle = 2\lambda \mathtt{E}_n \vert n,\ell \rangle, \quad H_{\ell} =-\frac{d^2}{dr^2} + \frac{\ell (\ell+1)}{r^2} +\lambda^2 r^2,
\label{eigen2}
\ee
where the eigenvectors $\vert n,\ell \rangle$ form an orthonormal set
\be
\delta_{nn'} \delta_{\ell \ell'} = \langle n,\ell \vert n',\ell'\rangle= \int_0^{\infty} \overline u_{n\ell}(r) u_{n' \ell}(r) dr \int_0^{\pi} \overline\Theta_{\ell, 0}(\theta) \Theta_{\ell', 0}(\theta) \sin \theta d\theta.
\label{inner2b}
\ee
Finally, the transformation from the second order linear differential equation (\ref{eigen}) to the confluent hypergeometric one \cite{Flu71,Neg00,Ros03} leads to the general form of the $u$-fnctions ($\gamma$ and $\delta$ are arbitrary constants):
\be
\begin{array}{rl}
u_{n\ell}(r) = &  r^{\ell +1} e^{-\lambda r^2/2} \displaystyle\left [ \gamma {}_1F_1 \left(\frac{\ell}{2} +\frac34 -\frac{\mathtt{E}_n}{2}, \ell + \frac32, \lambda r^2 \right) \right.\\[3ex]
& \qquad + \displaystyle\left. \delta r^{-(2\ell+1)} {}_1F_1 \left(-\frac{\ell}{2} +\frac14 -\frac{\mathtt{E}_n}{2}, -\ell + \frac12, \lambda r^2
\right)\right].
\end{array}
\label{ugral}
\ee
Finding regular $u$-functions produces the discreteness of the energy eigenvalue. For instance, taking $\delta=0$ and
\be
\frac{\ell}{2} +\frac34 -\frac{\mathtt{E}_n}{2} =-s \quad \Rightarrow \quad \mathtt{E}_n = 2s + \ell + \frac32 \equiv n+\frac32, \quad s,\ell, n =0,1,\ldots
\label{eigengral}
\ee
we arrive at the well known expression of the normalized solutions
\be
u_{n\ell}(r)=\left[ \frac{2 \Gamma(s+1)}{\Gamma(s+\ell +\frac32)}\right]^{1/2} r^{\ell +1} e^{-\lambda r^2/2} L^{(\ell +1/2)}_s (\lambda r^2),
\nonumber
\ee
where we have used the relationship between the confluent hypergeometric functions ${}_1F_1(a,c,z)$ and the associated Laguerre Polynomials $L_n^{(\gamma)}(z)$, see \cite{Olv10}.

\section*{Acknowledgment}

The support of CONACyT and SIP project 20160527 are acknowledged.


\end{document}